\documentclass[review]{elsarticle}
\usepackage{hyperref}
\usepackage{amsmath,amssymb,amsfonts}
\usepackage{algorithmic}
\usepackage{graphicx}
\usepackage{textcomp}
\usepackage{multirow}
\usepackage{subfig} 
\usepackage{booktabs}
\usepackage{adjustbox}

\hypersetup{colorlinks=false}

\hypersetup{
	colorlinks=false,
	pdfborder={0 0 0},
}

\begin{document}
	
	\begin{frontmatter}
		\title{FaultFace: Deep Convolutional Generative Adversarial Network (DCGAN) based Ball-Bearing Failure Detection Method}
		\author{Jairo Viola\fnref{myfootnote}}
		\author{YangQuan Chen\fnref{myfootnote}\corref{mycorrespondingauthor}}
		\author{and Jing Wang$^2$}
		\cortext[mycorrespondingauthor]{Corresponding author}
		\address{$^1$University of California, Merced, USA}
		\address{$^2$Beijing University of Chemical Technology, China}
		
		\begin{abstract}
			Failure detection is employed in the industry to improve system performance and reduce costs due to unexpected malfunction events. So, a good dataset of the system is desirable for designing an automated failure detection system.  However, industrial process datasets are unbalanced and contain little information about failure behavior due to the uniqueness of these events and the high cost for running the system just to get information about the undesired behaviors. For this reason, performing correct training and validation of automated failure detection methods is challenging. This paper proposes a methodology called FaultFace for failure detection on Ball-Bearing joints for rotational shafts using deep learning techniques to create balanced datasets. The FaultFace methodology uses 2D representations of vibration signals denominated faceportraits obtained by time-frequency transformation techniques. From the obtained faceportraits, a Deep Convolutional Generative Adversarial Network is employed to produce new faceportraits of the nominal and failure behaviors to get a balanced dataset. A Convolutional Neural Network is trained for fault detection employing the balanced dataset. The FaultFace methodology is compared with other deep learning techniques to evaluate its performance in for fault detection with unbalanced datasets. Obtained results show that FaultFace methodology has a good performance for failure detection for unbalanced datasets.
		\end{abstract}
	\end{frontmatter}

	\section{Introduction}
	In control engineering, failure detection is the branch concerned on monitoring a system, identify the possible failures, and notifying its kind and location using only the available input and output data streams of the system. So, it makes possible detecting not only the system failures but also discovering hidden behavior patterns, which are reflected in plant stops that generate productivity and money losses for the companies. Also, failure detection is a challenging task for different reasons like the system complexity, the required prediction speed response, the size, and consistency of the dataset, or the number of performance indices evaluated. 
	In the literature, there are several applications of machine learning and deep learning techniques for failure detection of industrial processes. In \cite{b1}, a support vector machine (SMV) is employed to detect failures inside a wireless sensors network due to damages in the devices or faults in the communication. On the other hand, \cite{b2} shows the use of unsupervised K-means algorithm to detect failures on 3D stacked integrated circuits. In \cite{b3}, a distributed machine learning classification algorithm to detect attacks into the power grid is shown, which use the K-means algorithm, SVM, decision tree, among other methods. Another application on semiconductors failure detection is given by \cite{b4}, where an assessment of different Machine learning models is performed to detect several types of failures during the wafer manufacturing process. Also, \cite{b5} presents a failure detection algorithm that employs logistic regression models to detect failures due to mechanical component fatigue.  
	On the other hand, \cite{b6} shows a prognosis method for shackles employing logistic regression to determine the decision boundaries for each failure. In the case of failure detection on robotic systems, \cite{b7} shows a comparison between classic machine learning, statistical procedures, and the hybrid boosted gradient method, which is an improvement of the logistic regression. There is also an application of machine learning techniques for failure detection on directional drilling of oil wells \cite{b8}, where the training process was performed using significant historical data from more than 80 oil wells for training a boosted gradient algorithm. Besides, machine learning can also be employed for Cyber-Physical Systems. In \cite{b9}, the random forest method is used to perform disturbance detection on a smart grid system. Also, \cite{b10} present a survey of various machine learning algorithms like SVM, Logistic Regression, and random forest for failure detection on the Internet of Things (IoT) sensor networks. As can be observed, the applications presented on \cite{b1} - \cite{b10} employs time series analysis, machine learning, or deep learning methods for training the classifiers and perform the failure detection. Notice that these applications have a good quality dataset, allowing a correct training of the failure detection algorithms. 
	
	However, on industrial processes, there is not always available a balanced, complete, or consistent dataset related with the failure behavior due to the longer time required to run a complete cycle of the process. Likewise, the cost and risk of running a process to get data from a failure behavior may produce more significant damages in the physical system. For this reason, the training of classifiers for industrial process sometimes relies totally on simulated data. For example, \cite{b11} shows an application where a machine learning algorithm is employed for early failure detection on CNC machines, which is trained using an identified state-space model of the system to generate the failure and nominal data of the machine. Also, \cite{b12}, employs a simulation model of an electric car power drivers to train a machine learning model for failure detection based on an artificial neural network. The main challenge for this approach is that a representative model of the system is not always available for training a machine learning model accurately.
	
	So that, fault detection in industrial processes with unbalanced datasets is an active research topic, which combines machine and deep learning techniques for fault classification and additional feature mining over scare fault data. For example \cite{b35}, presents the use of bilayer Convolutional Neural Networks (CNN) for fault detection in chemical processes with unbalanced datasets, which is based on a exhaustive feature mining of the available data using wavelet packet decomposition. In \cite{b36}, a CNN network combined with an initial normalization kernel is employed for fault detection in bearing mechanisms, mining additional data with the CNN convolution layers. Also, \cite{b37} presents the use of fusion autoencoders for skewed, incomplete, unbalanced datasets, with several denoising and resampling stages for feature extraction applied to fault detection on bearing elements. Notice that these works relays on deep feature extraction to compensate the unbalanced and incomplete dataset in order to improve the fault detection accuracy.
	
	On the other hand, Generative Adversarial Networks (GAN) \cite{b13}, proposed by Goodfellow in 2014, expand the reaches of Artificial Intelligence (AI) allowing the creation of new datasets based on small amounts of available data. These generated data is not only closer to the original but also can produce images combining different features extracted from the original dataset. For this reason, there are many applications of the GAN networks for classification problems. For example, \cite{b14} shows the use of GAN networks for the artificial generation of synthetic data for training a detection model of Jellyfish swarms. In \cite{b15}, a multi-class spectral GAN network is employed for the classification of multispectral images. Also, in \cite{b16}, a Multiview GAN network is proposed for pearls classification, increasing the accuracy regarding classical methods. Likewise, \cite{b17} shows the application of GAN for medical images generation and classification for different body diseases. For failure detection on industrial processes, some reported works are using different GAN networks for dataset generation. In \cite{b18}, the fault diagnosis is performed for a planetary gearbox system using GAN networks and Stacked Denoising Autoencoders. Besides, \cite{b19} and \cite{b20} present unsupervised classification algorithms for rolling bearings in combination with GAN networks, which contains an unbalanced dataset. For all the GAN networks applications presented above, the feature extraction process is performed using algorithms like Autoencoders, external to the GAN network. 
	
	Nonetheless, there is a particular implementation of the GAN network known as Deep Convolutional Generative Adversarial Networks (DCGAN), which incorporate the automatic feature extraction layers for the images with the GAN network. Thus, all the feature extraction and training process is performed using only this network. There are some applications that use DCGAN networks for medical image generation \cite{b21} \cite{b22}, or  image augmentation \cite{b23}. However, for failure detection, there are few applications of the DCGAN reported like \cite{b24}, where DCGAN is employed failure detection on photovoltaic systems or \cite{b25} where is employed for intrusion detection.
	
	This paper presents a fault detection methodology called FaultFace, which is employed for the failure detection on ball-bearing joints for rotational shafts using DCGAN networks for dataset balancing. The system to be analyzed is the Case Western Reserve university benchmark \cite{b26}, which is employed to evaluate different ball bearing joints faults on a rotational shaft axis. 
	
	A face portrait of the vibration signals is obtained for the nominal and failure behaviors, which correspond to a time-frequency representation of each signal. Six different FacePortraits are obtained from the vibration data, using Continuous Wavelet Transformation (CWT) with Morse Wavelet \cite{b29}, Wavelet transformation with HAAR Wavelet \cite{b30}, Circular Matrix Reading (CMR) \cite{b27}, Toepliz matrix \cite{Toep}, Hankel matrix, and Gramian matrix \cite{LinAlg}.
	
	Considering that the ball bearing dataset is unbalanced and contains few samples of nominal and failure cases, the DCGAN network is employed to generate new face portraits for the nominal and failure cases. Then, the balanced dataset generated by the DCGAN is used to train a Convolutional Neural Network (CNN) that perform the failure detection task. The structural similarity index (SSIM) is employed to measure the quality of the new dataset generated using the DCGAN network. Also, another balanced dataset is produced using a GAN network to compare not only the performance of the DCGAN network but also the overall performance of the faultFace methodology. The obtained results of the faultFace methodology are evaluated using the confusion matrix for the DCGAN and GAN datasets. The faultFace methodology is compared with a support vector machine (SVM) with Autoencoder and a Long Short Term Memory network (LSTM). Likewise, it is compared with other reported classification methods employed for the CWRU ball-bearing dataset.
	
	The main contribution of this paper is presenting the FaultFace methodology, which leverage the deep learning techniques like CNN and DCGAN usually employed for face recognition in failure detection for industrial processes with unbalanced datasets. Also, a comparison between the FaultFace methodology with other failure detection methods is performed to asses the capabilities of FaultFace to improve fault detection given balanced dataset.
	
	The structure of this paper is as follows. Section II presents the DCGAN and CNN networks employed for fault detection. Section III presents the ball-bearing benchmark system and the description of the nominal and failure behaviors of the system. Section IV introduces the faultFace methodology which involves the procedures used for facePortraits generation, the training of the DCGAN network for dataset balancing, the CNN training based on the new face portraits produced by DCGAN as well as the performance assessment of the methodology using the confusion matrix as well as a quality evaluation of the generated balanced dataset using the DCGAN network. Section V shows a variant of the faultFace methodology using the GAN network for dataset balancing instead of the DCGAN network as well as the performance comparison between both approaches. Section VI presents a comparison of the faultFace methodology with other proposed methodologies for failure detection of this system  including LSTM and SVM with Autoencoder. Finally, conclusions and future works are presented.
	
	\section{Deep learning tools for failure classification}
	
	\subsection{Generative Adversarial Networks (GAN)}
	According to \cite{b13}, a Generative Adversarial Network (GAN) is a deep learning model based on two independent neural networks called generator (G) and discriminator (D), which are involved in a competition. The generator (G) network creates a new probability distribution $P_G(x)$ based on a prior defined probability distribution $P(x)$, which can be considered as a black box. On the other hand, the discriminator (D) network determines the difference between the $P_G(x)$ and $P(x)$. Once the discriminator cannot distinguish between $P_G(x)$ and $P(x)$, it means that the generator learns the black-box behavior of $P(x)$. Notice that G and D are trained simultaneously in order to improve the estimation of $P_G(x)$ as well as the differentiation of $P(x)$ against $P_G(x)$. So that, the GAN network can be defined as a minimax optimization problem as given by (1), where $x\sim P_{data}(x)$ is the data from the original distribution $P_(x)$ and $z\sim P_z$ is the data from the distribution generated by G.
	\begin{eqnarray}
	\min_G \max_D V(D,G)= E_{x\sim P_{data}(x)}[Log D(x)]+E_{z\sim P_z}[log(1-D(G(z)))].
	\end{eqnarray}
	\par
	From (1), the GAN network tries to maximize the probability $log(D(G(z)))$ of an accurate classification by D, while simultaneously trying to minimize the error on G by $log(1-D(G(z)))$. A block representation of the GAN network is presented in Fig.1. As can be observed, the generator network is feed with a random noise distribution to generate $P_G(x)$, which feed the discriminator network to determine whether the synthetic data produced by the generator is real or fake, and based on that result perform the training of the generator and the discriminator again. The minibatch stochastic gradient descent is employed as a training algorithm for the GAN network \cite{b13,b27}. For the GAN network, the optimal training point is reached when $P(x)=P_G(x)$. Besides, the training process of G and D is performed simultaneously, reducing K times the gradient for training D and once for G, considering that the time for training D is higher than G.
	\begin{figure}
		\centering
		\includegraphics[width=8cm,height=2cm]{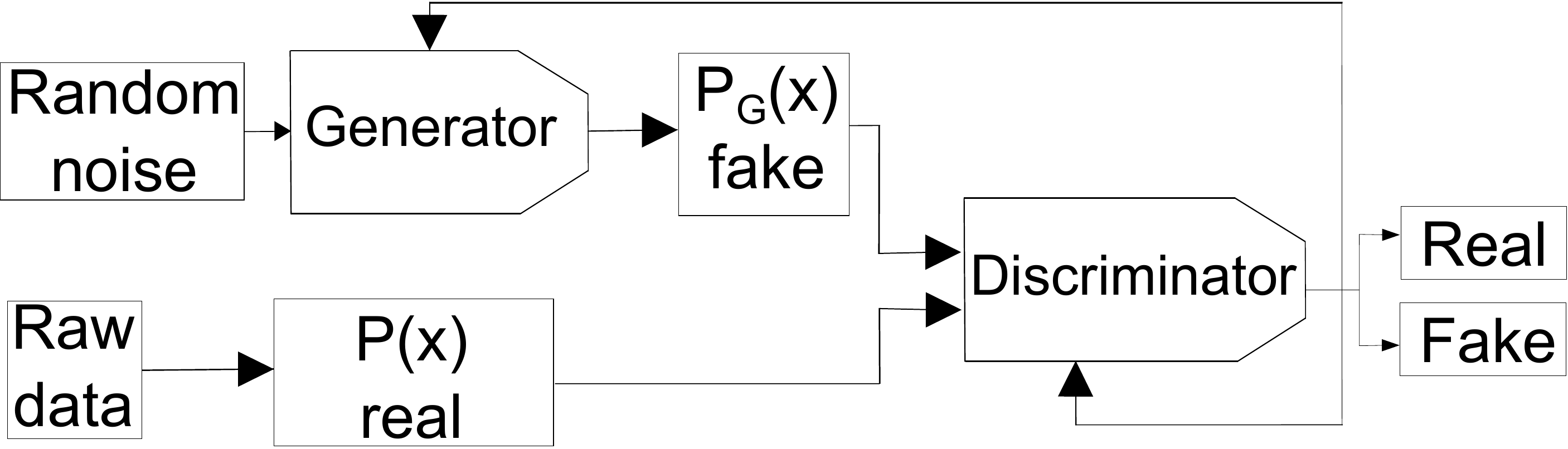}
		\caption{Block diagram of GAN network }
		\label{fig1}
	\end{figure}
	
	\subsection{DCGAN network}
	The deep convolutional GAN network (DCGAN) is a variation of the GAN network, where the generator and discriminator multilayer perceptron neural networks are replaced by a convolutional neural network to exploit its image processing capabilities. According to \cite{b28}, the CNN networks employed on the DCGAN network architecture should have some specific features to ensure a stable training process of the generator and discriminator. The first one is replacing the pooling layers with strided convolutions for the discriminator, and fractional-strided convolutions for the generator. The second one is eliminating full layers connections in the hidden layers of the generator and discriminator, just leaving the output layer fully-connected. The third one is to apply batch normalization to all the hidden layers expect by the input and output layer on the generator and discriminator, ensuring zero mean and unit variance. The fourth one is to use the ReLU activation function for the input and hidden layers, and Tanh activation function for the output of the generator to accelerate the training process. Finally, the LeakyReLU activation function is recommended for all the layers on the discriminator.
	\begin{figure}[h]
		\centering
		\includegraphics[width=8cm,height=4cm]{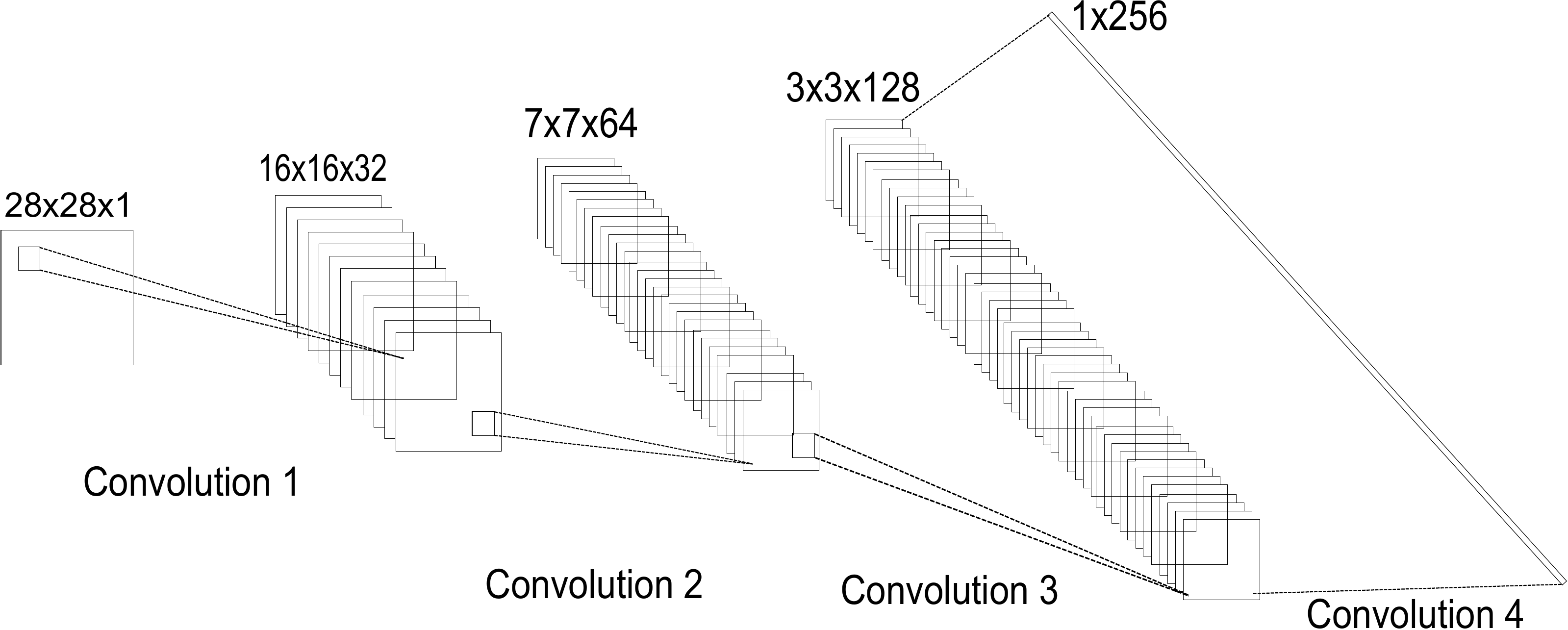}
		\caption{Discriminator CNN \cite{b10} }
		\label{fig2}
	\end{figure}
	The discriminator network employs the standard structure of a CNN presented in Fig.2.  As can be observed, the discriminator CNN has an input layer of 28x28. Also, three hidden layers are employed with LeakReLU as the activation function. Finally, the output layer has a dimension of 256x1, which is fully connected with a Sigmoid activation function for the real and fake data classification. The kernel size for the CNN is 3x3 in all its layers with striding of 2 for all the hidden layers except by the output with striding of 1.
	\par
	Besides, the generator network structure differs from the discriminator CNN, as shown in Fig.3. As can be observed, the generator CNN works perform the inverse CNN process. Initially, the sample random noise goes from a minibatch of Gaussian Random noise samples projected into a bigger feature space. After that, a 3x3 convolutional filter is applied, and the result is upsampled using a striding factor of 2, resulting in a higher-dimensional space. Thus, after some convolution layers, the generator returns a 2D image representation of the data. In this paper, the minibatch has an initial size of 100 samples, which is projected into a 128 feature dimensional space representation to apply three hidden convolutional layers with an upsampling factor of 2 that generate a 28x28 pixels 2D grayscale image in the output layer. That will be compared with the discriminator to perform the DCGAN network training. 
	\begin{figure}[h]
		\centering
		\includegraphics[width=8cm,height=3.5cm]{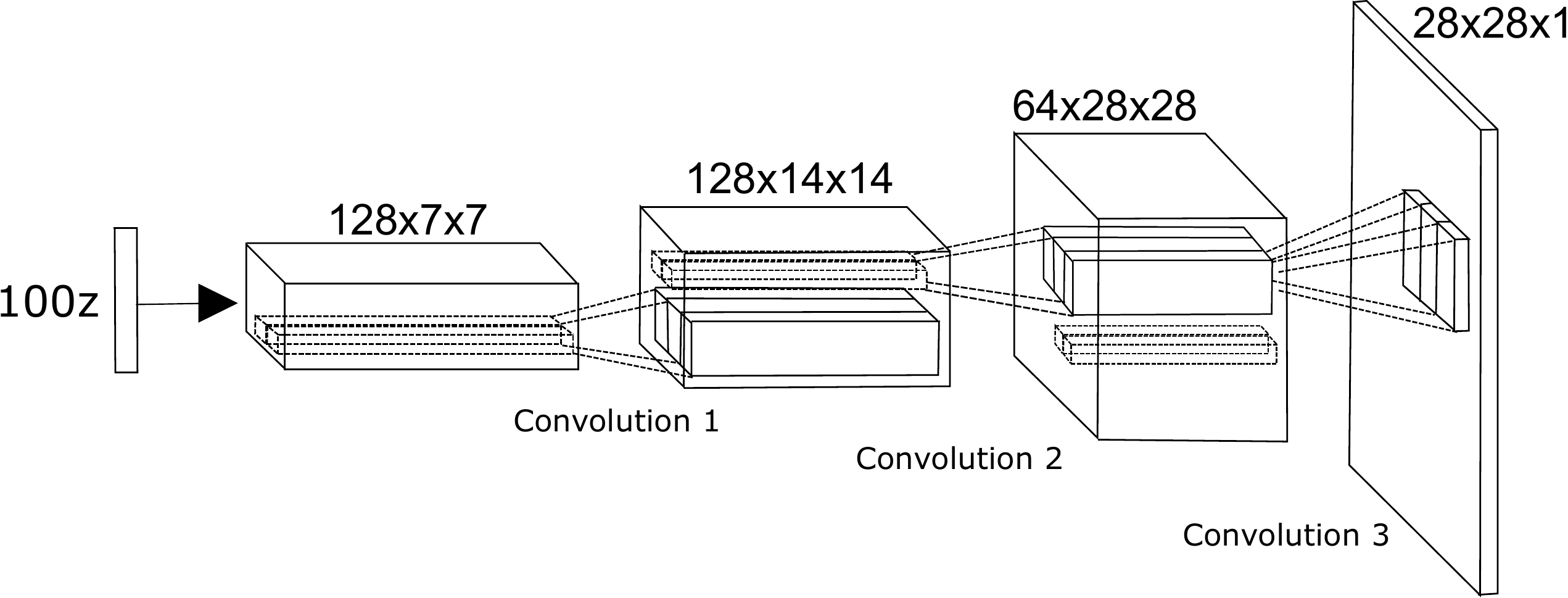}
		\caption{Generator inverse CNN \cite{b27}}
		\label{fig3}
	\end{figure}
	
	\section{Study case: Ball-Bearing benchmark system}
	The Ball-Bearing benchmark system from Case Western Reserve University  \cite{b26} and Rockwell automation were selected for testing the FaultFace method.  The benchmark system is presented in Fig.4. It is composed of two DC motors of $2Hp$ running at 1700 RPM which rotational shafts are joined using a ball-bearing coupling. This reference system is designed for testing different ball-bearing couplings diameters as well as inducing failures on the couplings using electrical pulses. For this system, the diagnosis signal is the axis vibration measured with accelerometers for different nominal and failure operating conditions.
	\begin{figure}
		\centering
		\includegraphics[width=0.5\textwidth,height=0.15\textheight]{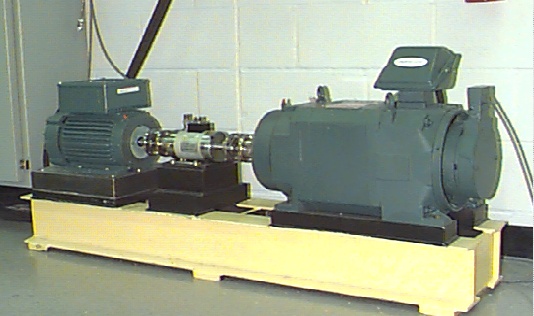}
		\caption{Ball-Bearing Benchmark system \cite{b26}}
		\label{fig4}
	\end{figure}
	\subsection{Ball-Bearing coupling failures}
	A ball-bearing coupling is presented in Fig.5. As can be observed, it is composed by an outer race (A), an inner race(B), the balls between the inner and outer race (C) to reduce the friction over the rotational shaft (D). According to \cite{b26}, different failures can be induced into the ball-bearing benchmark system. The first failure corresponds to damage on the inner race of the ball bearing, the second one is related to failures on the outer race due to the load position in the shaft (centered, opposite, orthogonal), and the third case is related to damages on the bearing balls. Table 1 summarize a set of possible failures for the benchmark system. As can be observed, the plant supports two different types of ball-bearing couplings denominated fan-end and drive-end with the possibility of generating different failure diameters.
	\begin{figure}
		\centering
		\includegraphics[width=0.45\textwidth,height=0.15\textheight]{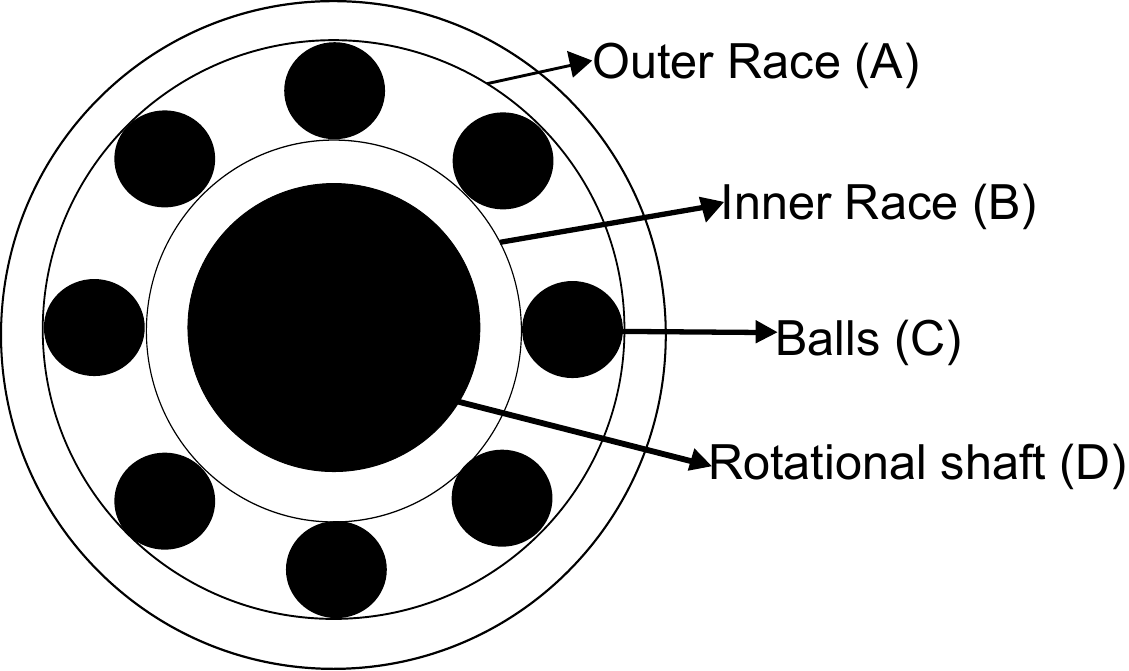}
		\caption{Ball-Bearing Coupling }
		\label{fig5}
	\end{figure}
	\begin{table}
		\caption{{Ball-Bearing benchmark system failures}}
		\begin{tabular}{|c|c|c|}
			
			\hline
			\textbf{Bearing type}                                                        & \textbf{Fault location}                                                                                                            & \textbf{Fault diameters}             \\ \hline
			\multirow{2}{*}{\begin{tabular}[c]{@{}c@{}}Fan end\\ Drive end\end{tabular}} & \multirow{2}{*}{\begin{tabular}[c]{@{}c@{}}Ball, Inner race, and Outer race load \\ ( Center, Opposite, Orthogonal)\end{tabular}} & \multirow{2}{*}{0.004, 0.014, 0.028} \\
			&                                                                                                                                    &                                      \\ \hline
		\end{tabular}
	\end{table}
	
	\subsection{Ball-Bearing dataset}
	The Ball-Bearing benchmark system is composed of 114 datasets of the rotational shaft vibration signal. Four datasets correspond to the nominal operation of the ball bearing coupling for fan end and drive end couplings. The remaining datasets are for the different failure behaviors of the system presented in Table.1. The data format is given as time series with sample rates of 12 kHz for the fan-end and 48 kHz for the drive-end Ball-Bearings. From the features presented above could be inferred that the Ball-Bearing Benchmark system is unbalanced with different sample rates. An example of nominal and failure behaviors time series are presented in Fig.6. It can be observed that the nominal and failure datasets were sampled by different times, and the failure vibration signals have a bigger amplitude than the nominal data for all the five failure cases. Therefore, the dataset should be balanced to obtain good performance from the failure classification technique. In this paper, the benchmark dataset is divided into six categories for classification and training purposes. The first one is denominated nominal data considering all the nominal datasets for different ball bearing types and sampling times. The other categories, corresponding to the failure cases are divided into the ball failure case, inner race case, outer race with centered load case, outer race with opposite load case, and outer race with orthogonal load case. 
	\begin{figure}
		\centering
		\subfloat[]{\includegraphics[width=0.3\textwidth,height=0.15\textheight]{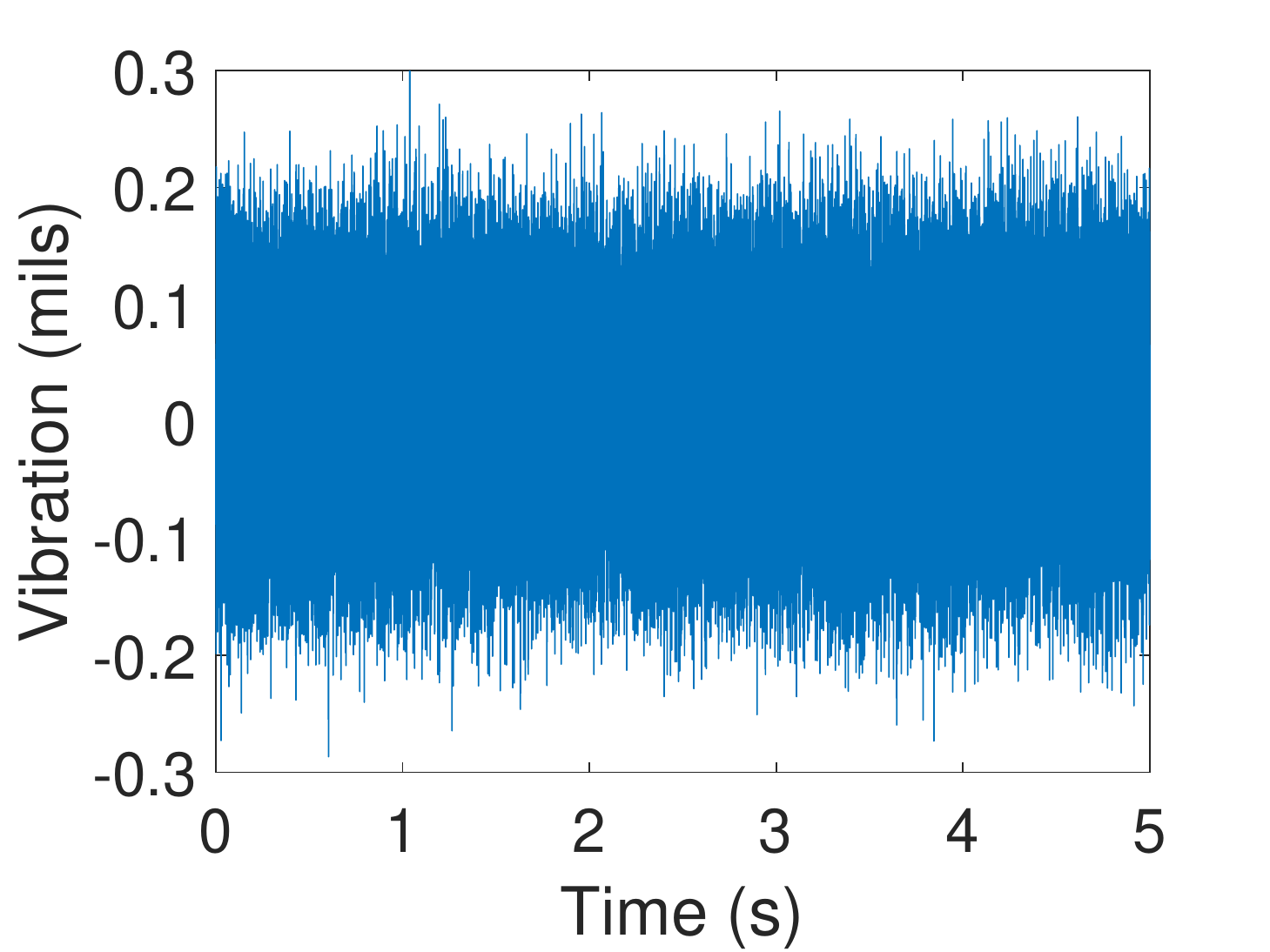}}
		\subfloat[]{\includegraphics[width=0.3\textwidth,height=0.15\textheight]{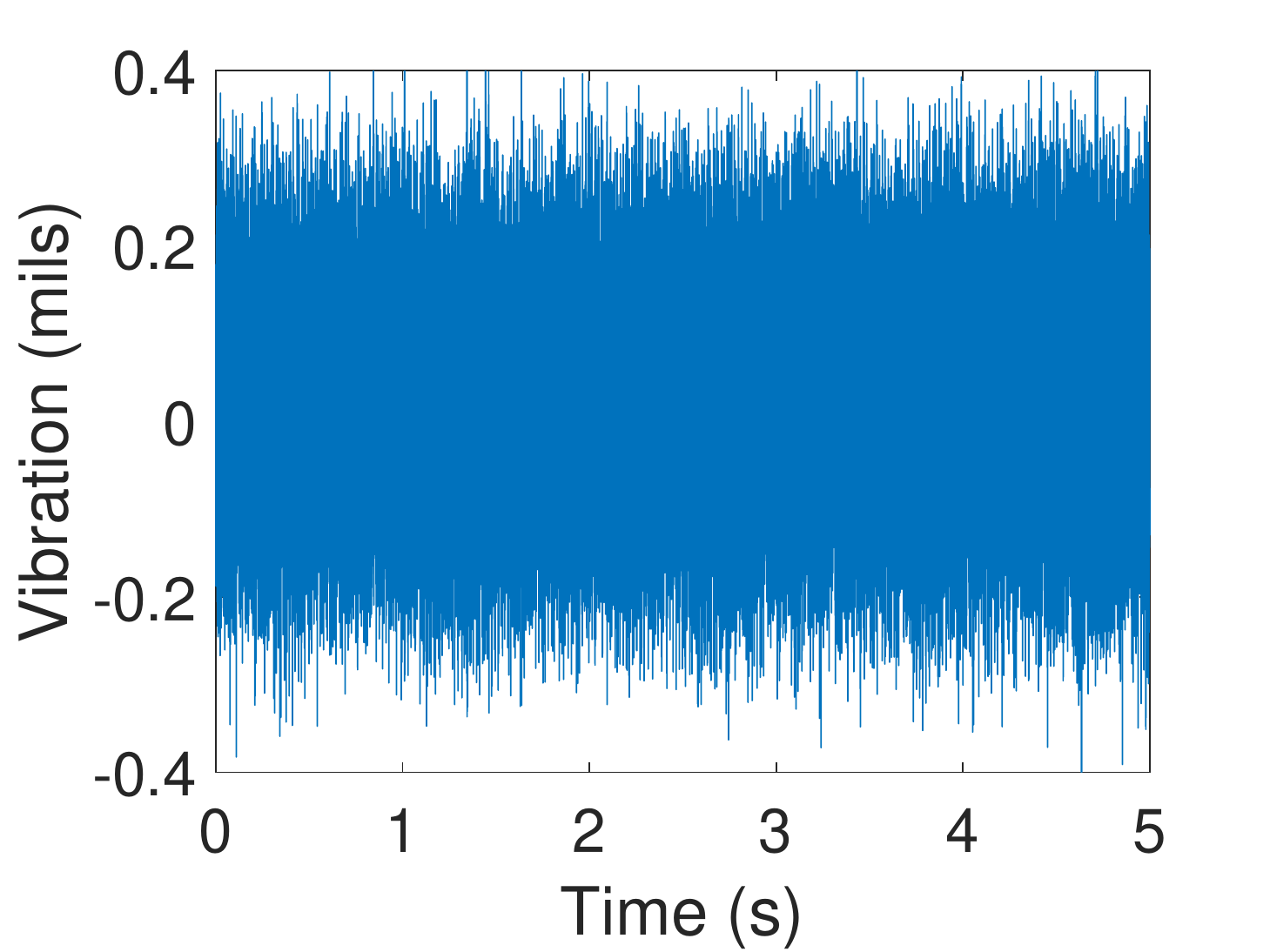}}
		\subfloat[]{\includegraphics[width=0.3\textwidth,height=0.15\textheight]{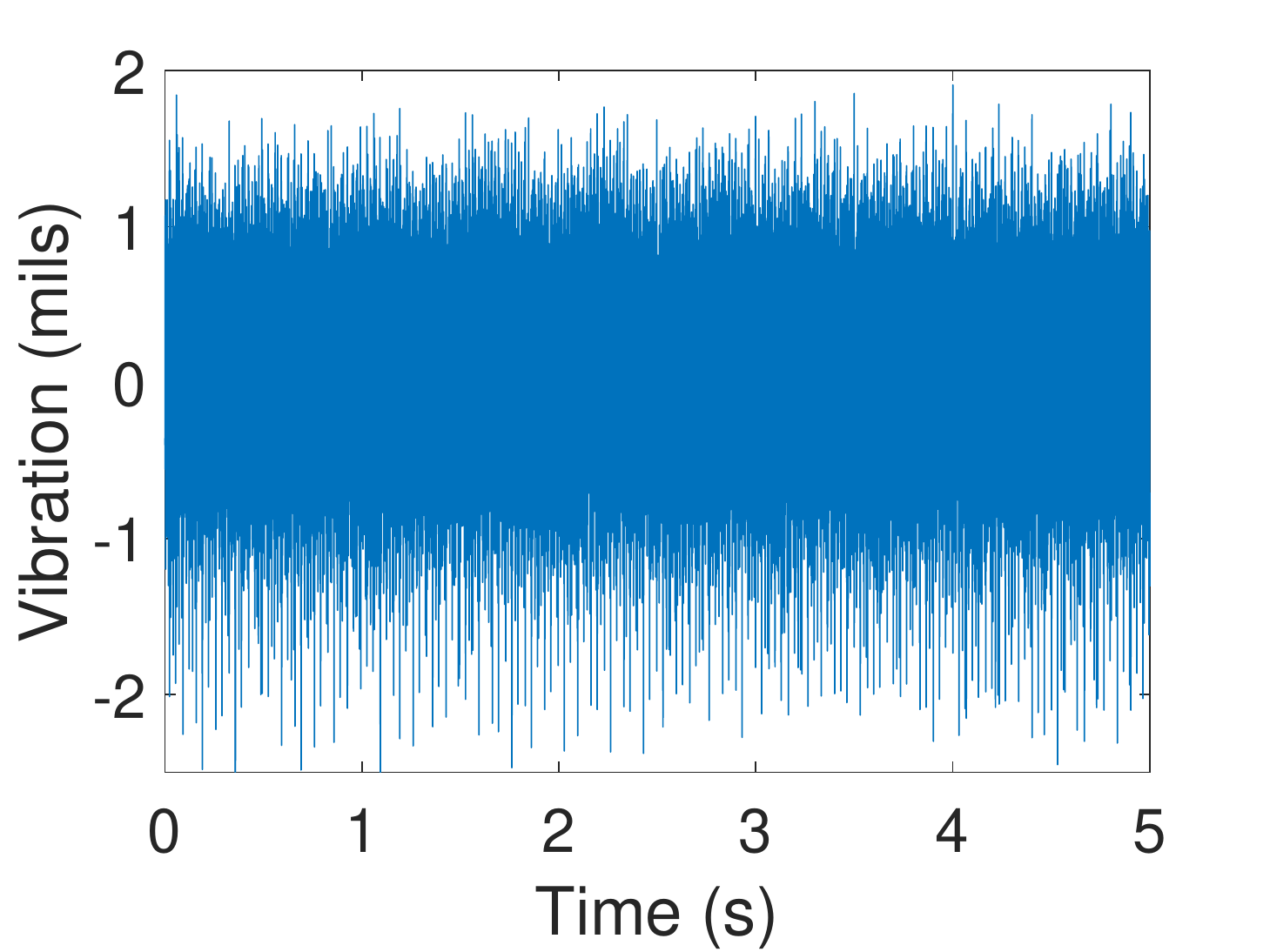}}\\
		\subfloat[]{\includegraphics[width=0.3\textwidth,height=0.15\textheight]{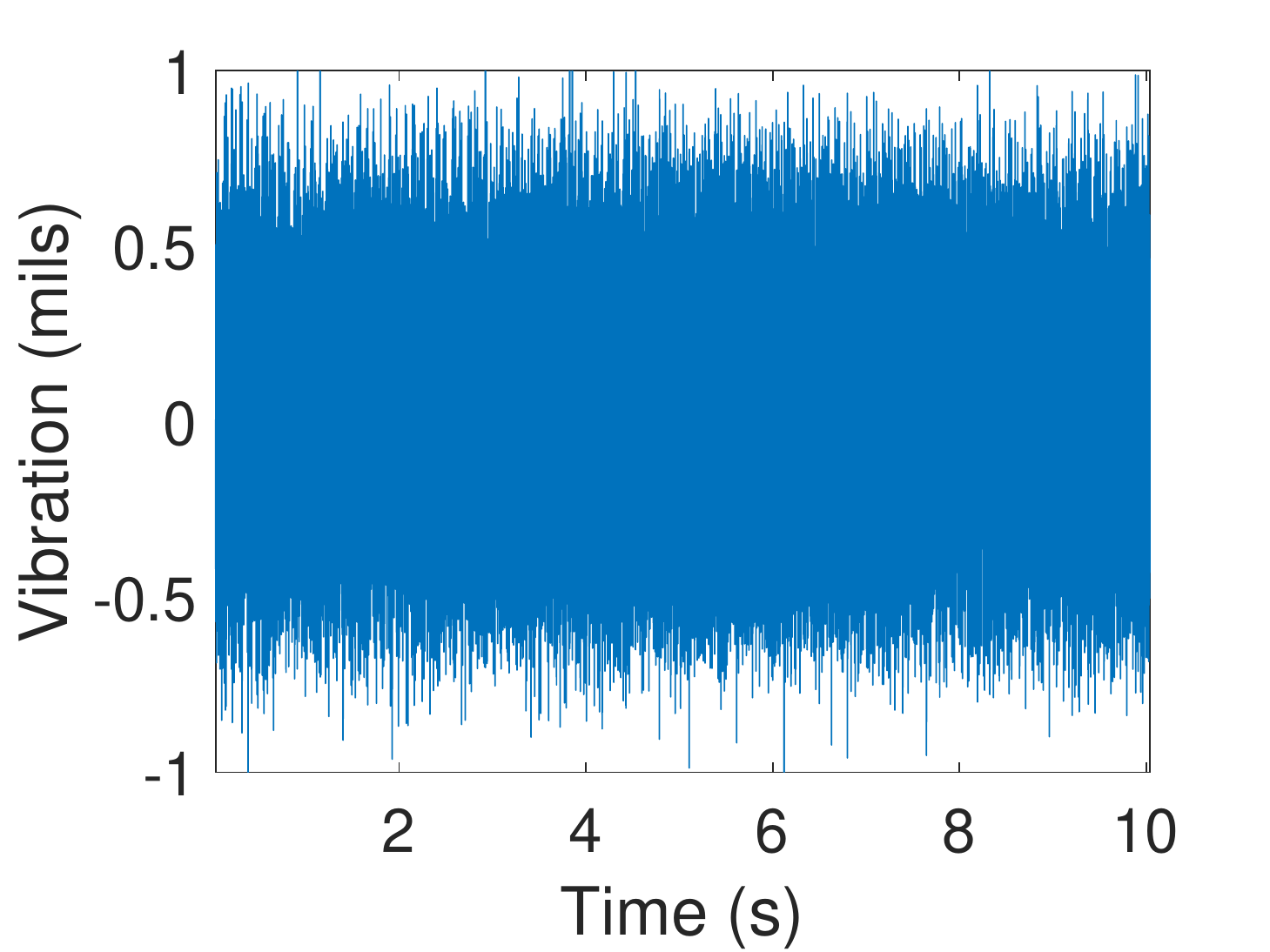}}
		\subfloat[]{\includegraphics[width=0.3\textwidth,height=0.15\textheight]{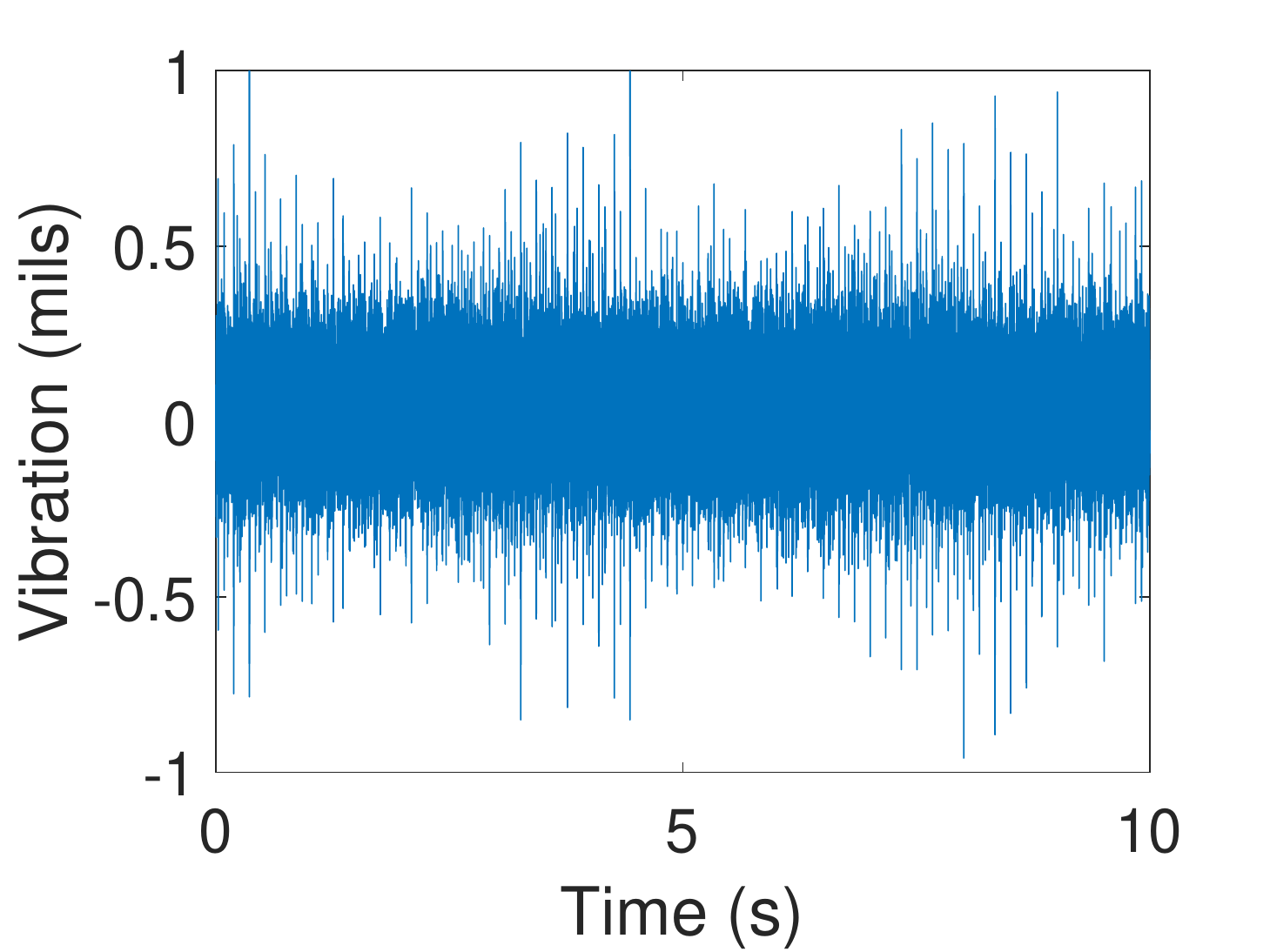}}
		\subfloat[]{\includegraphics[width=0.3\textwidth,height=0.15\textheight]{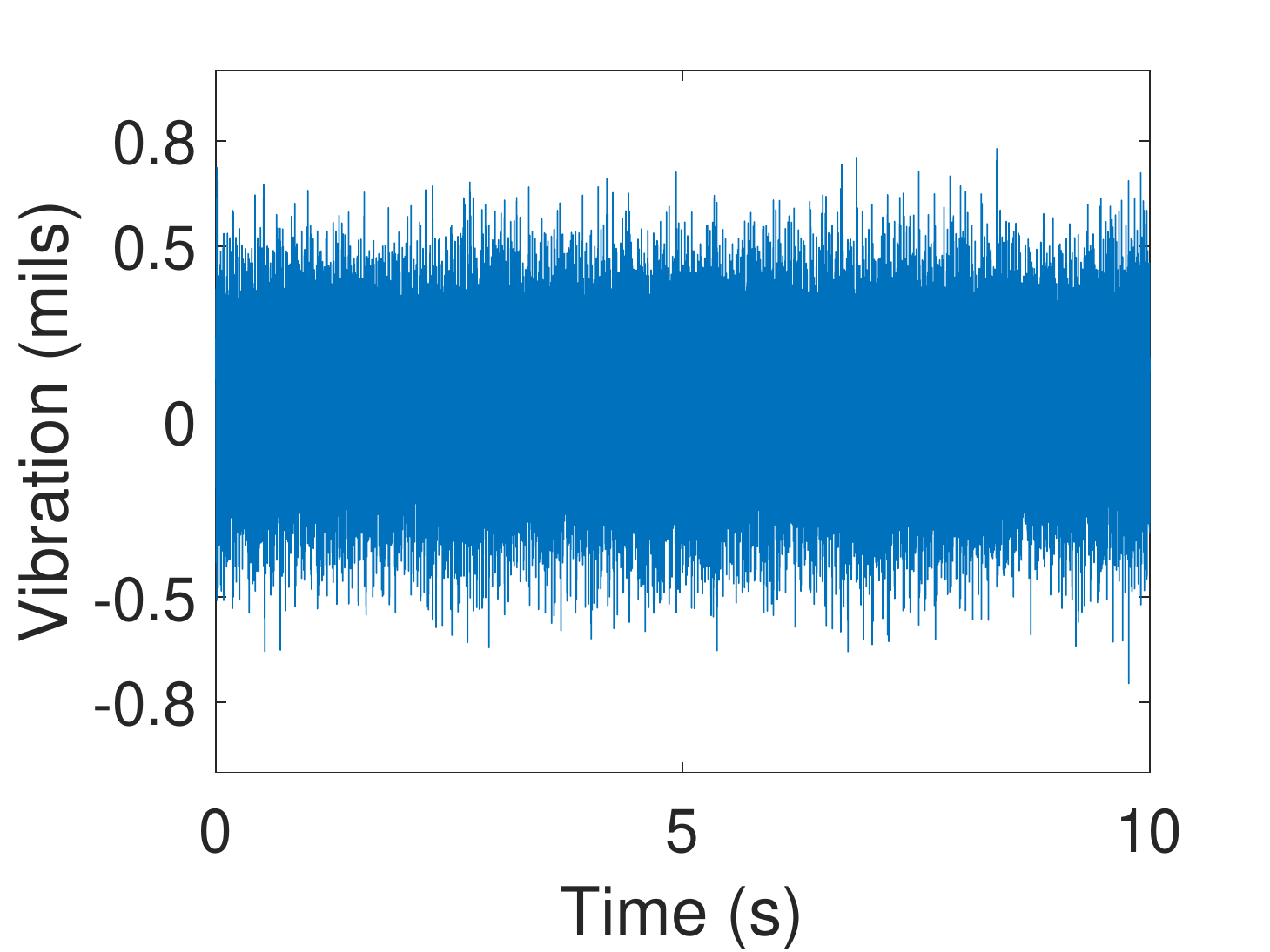}}	
		\caption[caption]{Vibration signals a) nominal b) Inner Race c)Ball d)Outer race: load center e)Outer race: load opposite and f)Outer race: load Orthogonal}
		\label{fig6}
	\end{figure}
	
	\section{FaultFace methodology}
	The block diagram representation of the FaultFace methodology is presented in Fig.7. Initially, the original unbalanced dataset of the ball-bearing nominal and failure behaviors is acquired. Then, the FacePortrait of the signals is determined. After that, the nominal and failure FacePortrits are introduced into the DCGAN network to generate new face portraits in order to balance each dataset. Next, using the new balanced datasets for nominal and failure behaviors generated from DCGAN, a Convolutional Neural Network is trained for failure detection. Finally, the obtained results are evaluated using the confusion matrix.
	\begin{figure}
		\centering
		\includegraphics[width=0.9\textwidth,height=0.35\textheight]{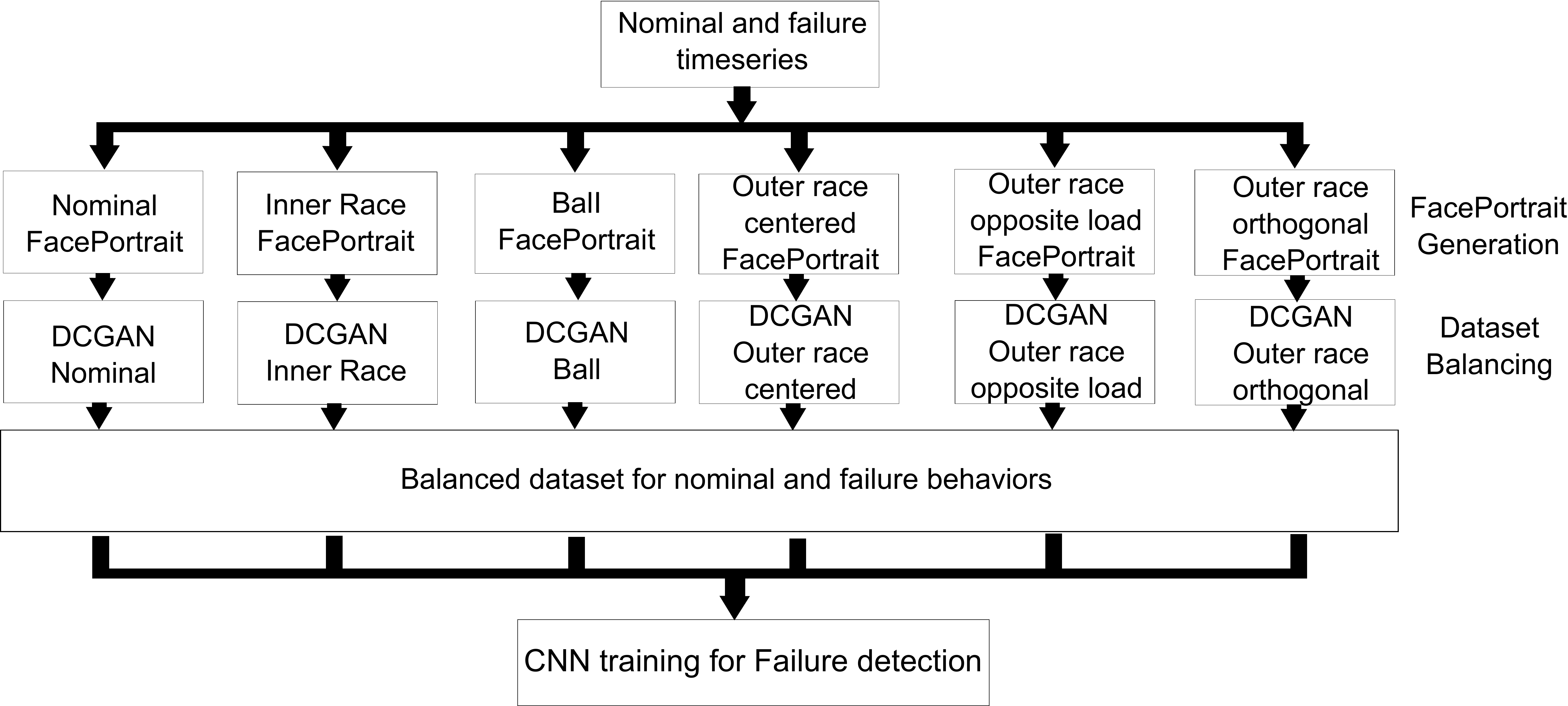}
		\caption{FaultFace methodology}
		\label{fig7}
	\end{figure}
	\subsection{Face portraits generation}
	The face portrait is a 2D image representation of a time series, which can be obtained employing time-frequency techniques. Six different face portraits representation for each signal ball-bearing vibration signal are obtained. The first one employs Continuous Wavelet Transformation (CWT) using the Morse wavelet \cite{b29}. The second one employs the Haar wavelet (HAAR) \cite{b30} instead of Morse wavelet. The third method employed is called Circular Matrix Reading (CMR) \cite{b27}. It consists of reading the time series, normalize regarding its maximum value and multiply each value of the time series by 255 to obtain a grayscale image of the time series, where each pixel represents a single value of the vibration signal. The fourth faceportrait uses a Toeplitz matrix transformation \cite{Toep}. It produces a symetric Toepliz matrix from the normalized vibration timeseries, where the elements along a diagonal have the same value. Likewise, the fifth faceportrait employs a Hankel transformation matrix \cite{LinAlg}. Unlike Toeplitz matrix, this transformation produce a symetric matrix where the antidiagonals elements are equal. The sixth faceportrait is generated using the Gram matrix $G$ \cite{LinAlg}, that is defined as all the possible inner products of $m$ vectors that conforms the set $V$. It is defined by $G=A^TA$, where A is a matrix with all the $m$ vectors of $V$ distributed as columns. In this paper, the $m$ columns for the matrix  $A$ were generated splitting the normalized vibration timeseries into equal length vectors. An example of the obtained face portraits for nominal and failure datasets is shown in Fig.8. As can be observed, all the vibration FacePotraits were transformed into a 28x28 pixels grayscale image that can be employed for training the DCGAN network for dataset balancing. Notice that each FacePortrait contains particular features that allow differentiating between nominal and failure behaviors. These features will be considered during CNN training in order to perform failure detection.
	\begin{figure}
		\centering
		\includegraphics[width=0.9\textwidth,height=0.3\textheight]{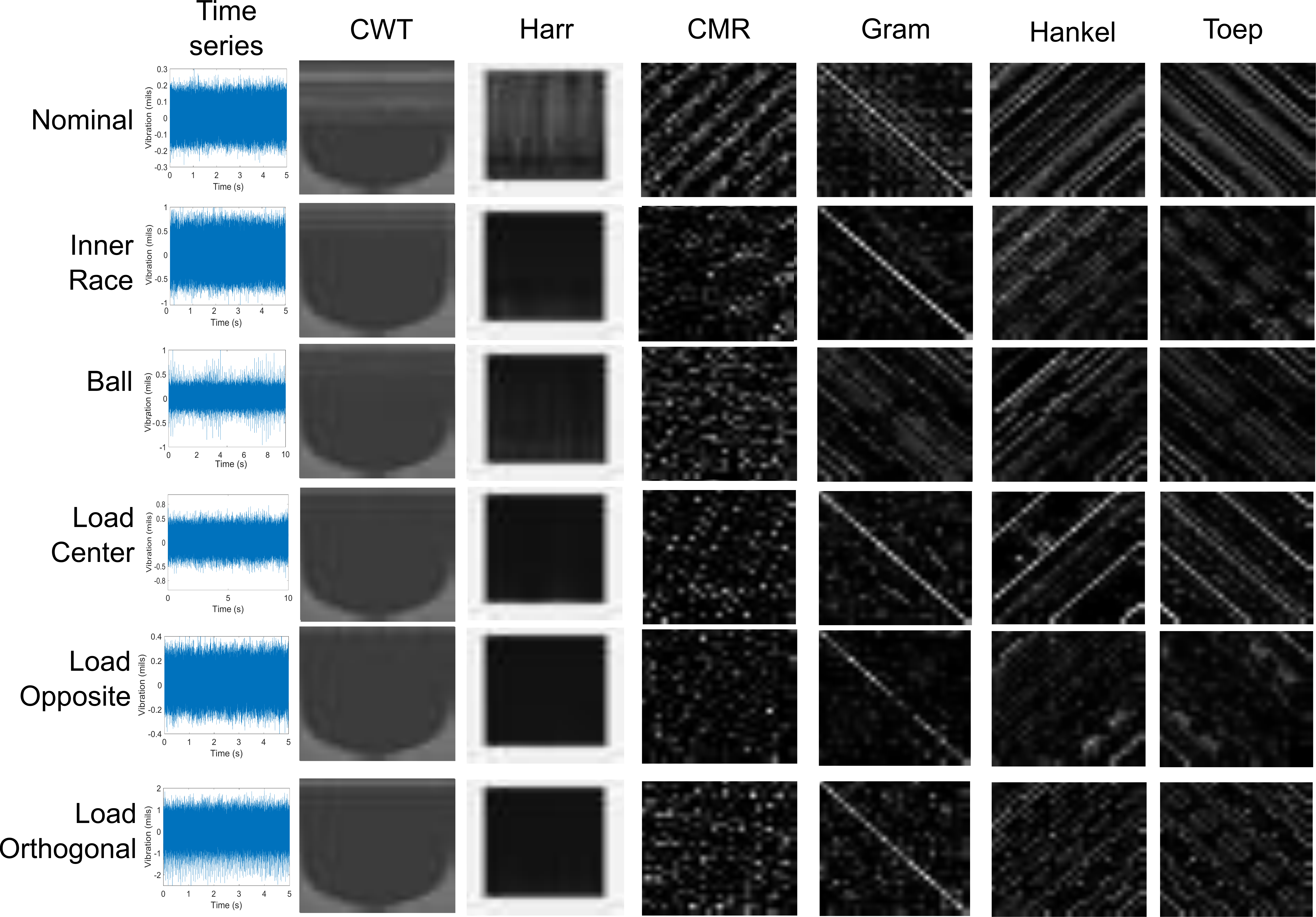}	
		\caption[caption]{Vibration signal and obtained FacePortraits for nominal behavior, Inner race, Ball, load center, load opposite, and orthogonal load failures}
		\label{fig8}
	\end{figure}
	
	\subsection{Dataset balancing using DCGAN Network}
\begin{figure}[h]
		\centering
		\includegraphics[width=1\textwidth,height=0.25\textheight]{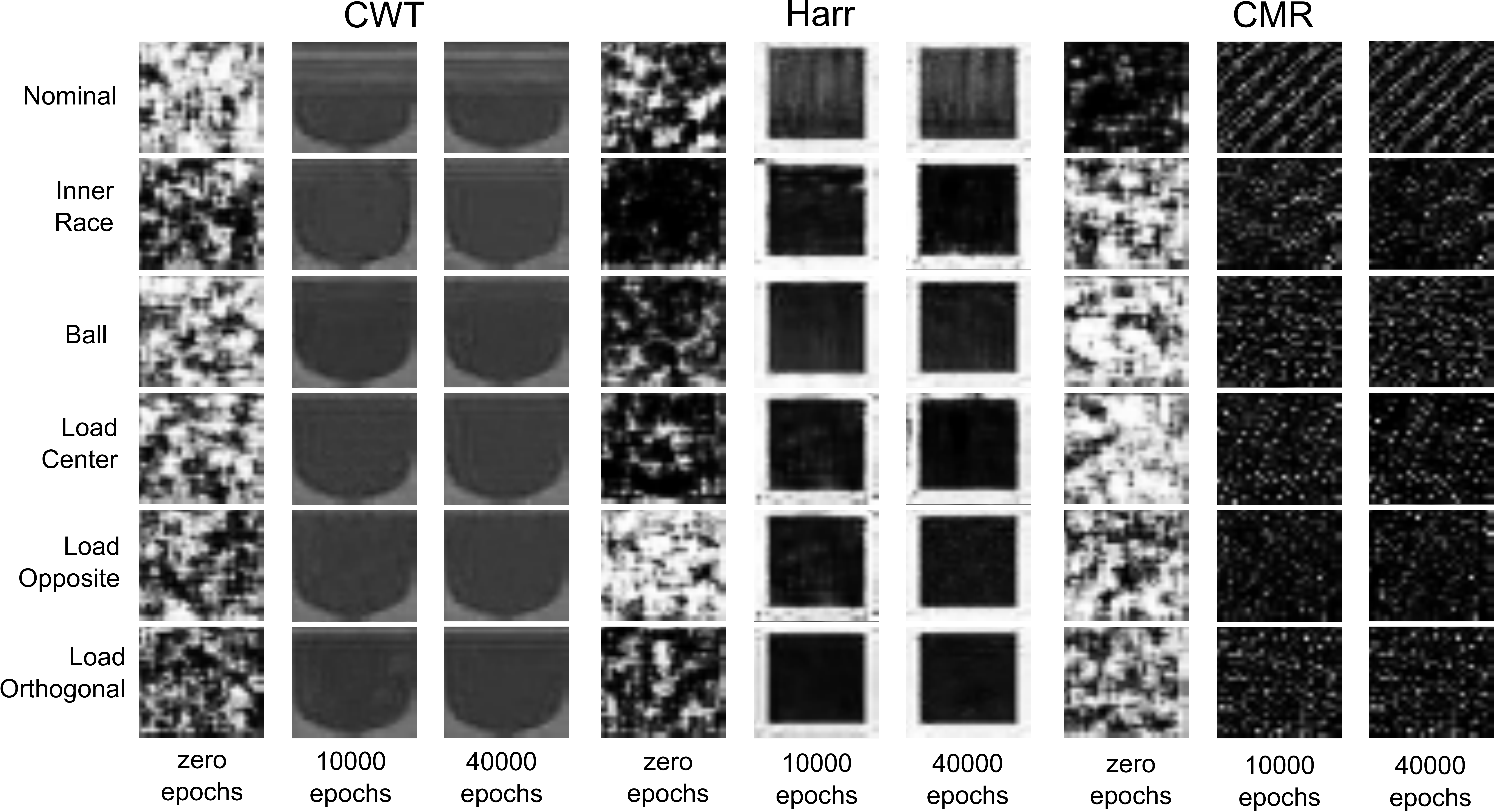}	
		\caption[caption]{DCGAN generated CWT, Harr, and CMR face portraits for nominal, Inner race, Ball, load center, load opposite, and load orthogonal at zero, 10000, and 40000 epochs}
		\label{fig9}
	\end{figure}

\begin{figure}[h]
		\centering
		\includegraphics[width=1\textwidth,height=0.25\textheight]{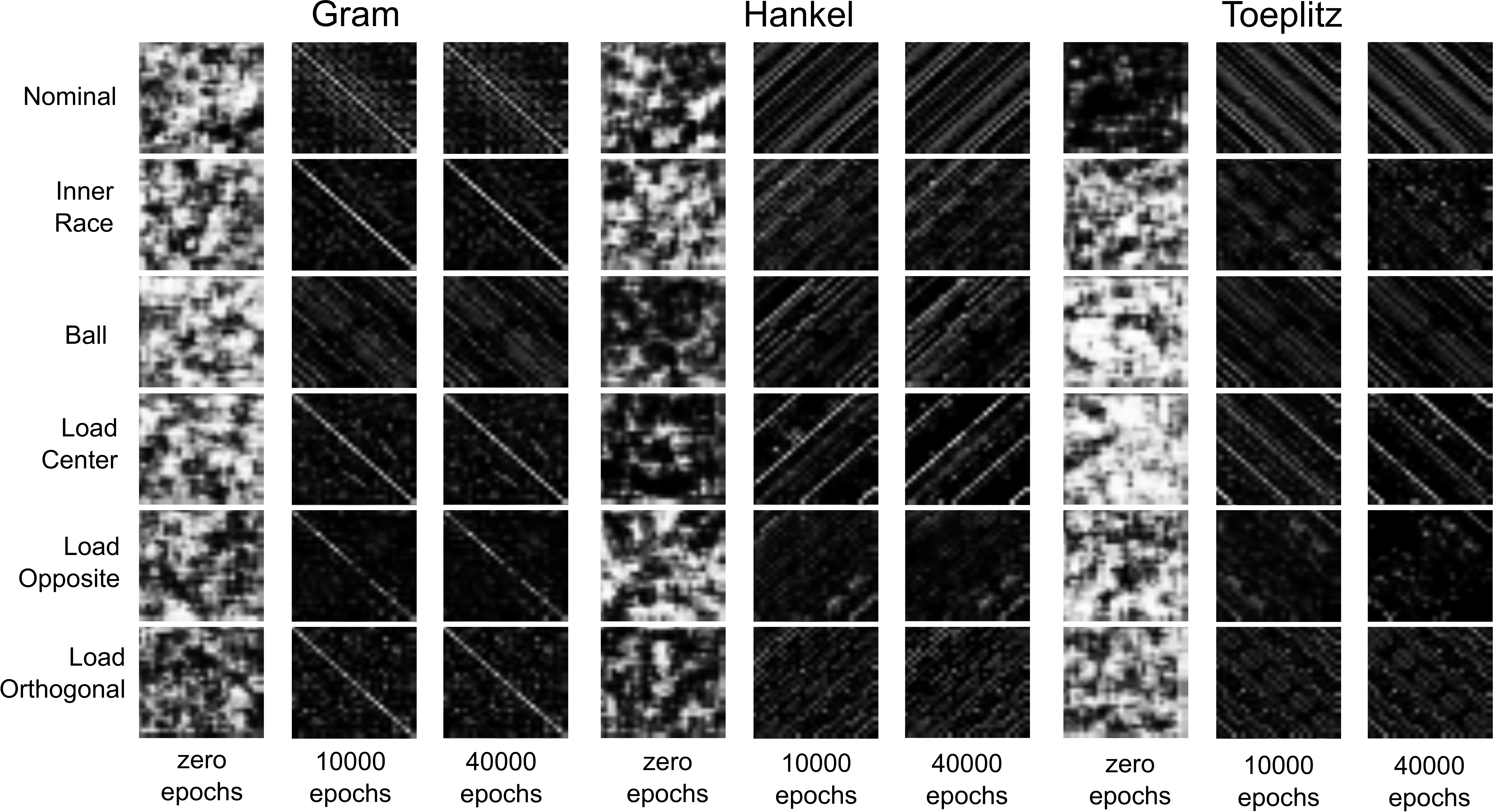}	
		\caption[caption]{DCGAN generated Gram, Hankel, and Toeplitz face portraits for nominal, Inner race, Ball, load center, load opposite, and load orthogonal at zero, 10000, and 40000 epochs }
		\label{fig10}
	\end{figure}

	The face portraits for nominal and failure behaviors shown in Fig.\ref{fig8} are introduced into a DCGAN network to produce a balanced dataset. In this paper, a individual DCGAN network was trained for the nominal behavior as well as for each fault case. For all the cases, each DCGAN networks were implemented in Tensorflow using the Keras framework and were trained with the minibatch stochastic gradient descent algorithm, using the Adam optimizer with a learning rate of 0.0001 for 40000 epochs. The results of the DCGAN network training for the CWT, Haar, and CMR faceportraits are shown in Fig.\ref{fig9}. Likewise, the Gram, Hankel, and Toeplitz faceportraits are shown in Fig.\ref{fig10}. As can be observed, the first epoch of the DCGAN generates an image that does not represent the face portrait and looks like random noise for all the cases. However, after 10000 epoch of training, the DCGAN networks begin to produce consistent face portraits, and after 40000 epochs, the result is similar to the original FacePortraits. Once the training process finishes, a balanced dataset is produced, which is composed of 1000 images of nominal behavior, and 1000 images for each failure behavior, it means a total of 6000 images. Notice that the original dataset only contains 114 time-series data, which only four represent the nominal behavior of the system.
	
	\subsection{DCGAN faceportrait quality assessment}
	A quantitative quality assessment of the balanced dataset produced by the DCGAN networks is performed to evaluate its accuracy for recreating the data distribution of the faceportraits. Thus, the structural similarity index (SSIM) is employed to measure the similarity of the generated faceportraits regarding to the original dataset. According to \cite{SSIM}, the SSIM is given by \eqref{ssim} for two images $x$, and $y$, where $\mu_x, \mu_y$ $\sigma^2_x, \sigma^2_y$, $sigma_{xy}$ correspond to the means, standard deviations and cross-covariance  of $x$ and $y$. Likewise,  $C_1,C_2,C_3$ are the regularization constants given by $C_1=(0.01L)^2$, $C_2=(0.03L)^2$, and $C_3=C_2/2$ with $L=255$ as the dynamic range for grayscale images.
	The SSIM index \eqref{ssim} returns a normalized value between [-1,1] where 1 represents a perfect matching between images $x$ and $y$.
	\begin{equation}
	SSIM(x,y)=\frac{(2\mu_x\mu_y+C_1)(2\sigma_{xy}+C_2)}{(\mu^2_x+\mu^2_y+C_1)(\sigma^2_x+\sigma^2_{y}+C_2)}
	\label{ssim}
	\end{equation}
	\par
\begin{figure}
		\centering
		\subfloat[]{\includegraphics[width=0.5\textwidth,height=0.3\textheight]{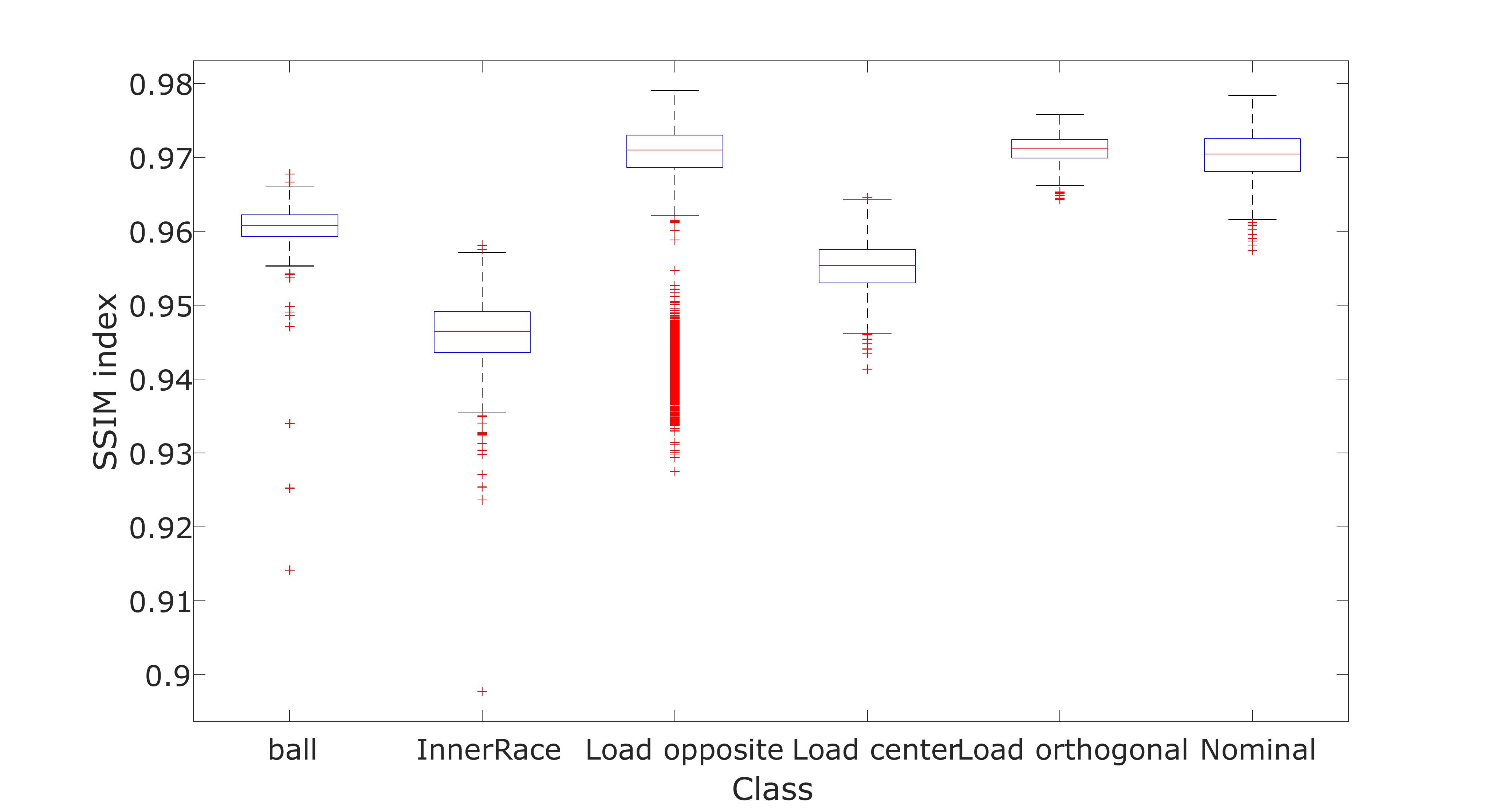}}
		\subfloat[]{\includegraphics[width=0.5\textwidth,height=0.3\textheight]{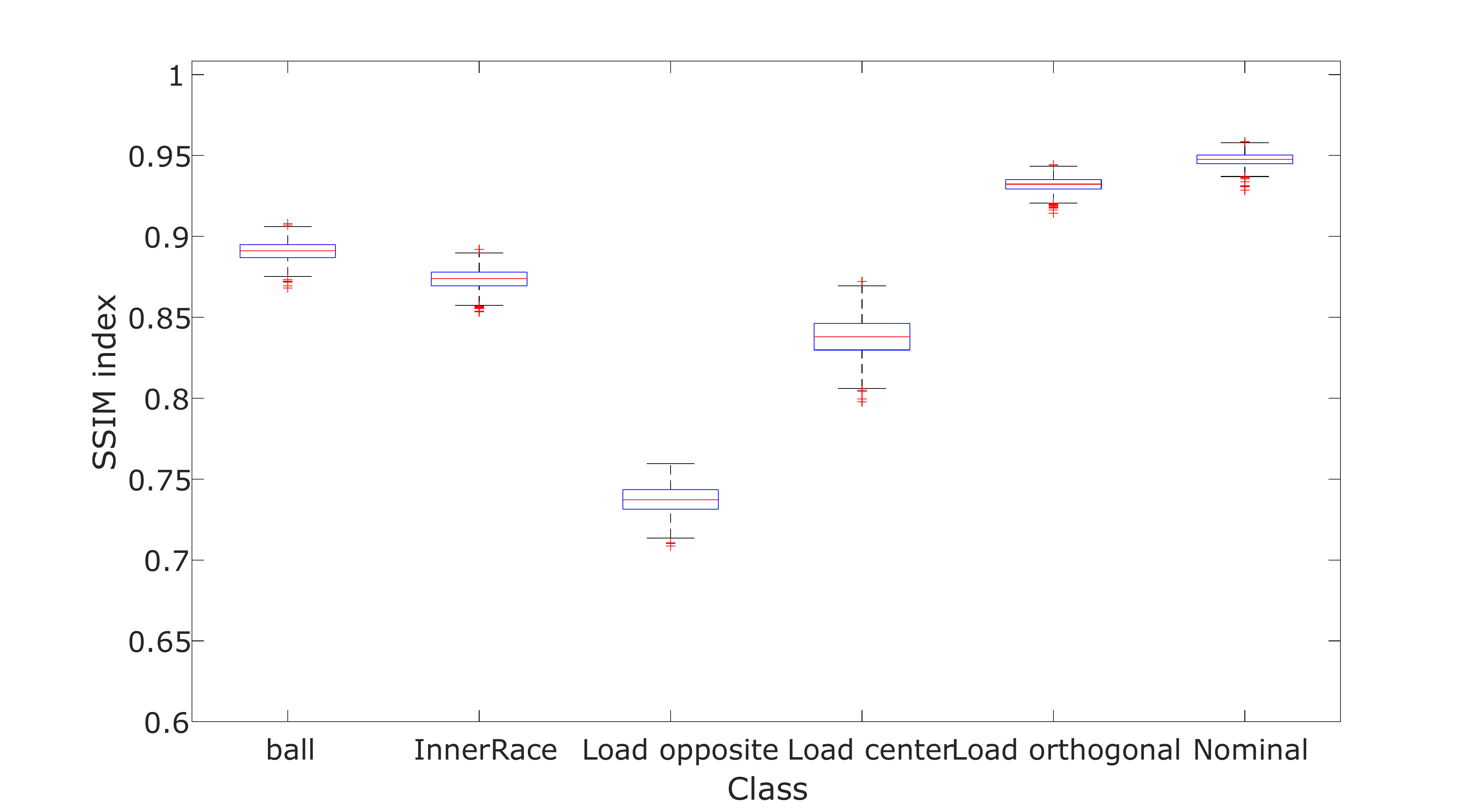}}
		\caption[caption]{SSIM index for quality assesment of the balanced dataset produced by the DCGAN network for the a) CWT and b)Hankel faceportraits}
		\label{dcganVal}
	\end{figure}

	In this paper, the SSIM index is calculated for each single image of the original faceportrait dataset with respect to each single image generated for the DCGAN network for each case and faceportrait in order to see the distribution of the generated faceportraits. As example,  Fig.\ref{dcganVal} shows a boxplot of the SSIM index calculated for the nominal CWT and Hankel faceportraits for the nominal and fault behaviors. 
	As can be observed, the mean value for the SSIM index for the CWT faceportrait is above of 94\% indicating a high similarity between the generated and the original dataset. Also, the deviation of the data is $\pm3\%$, which also indicates that the balanced dataset can improve the detection range of the faultFace methodology. In the case of Hankel faceportrait, the average SSIM index variates between 74\% to 95\%. In this case, the balanced dataset using Hankel faceportrait still performs a good representation of the system. 
	In addition, the data distribution is symmetric and follows a normal distribution, considering that the DCGAN network uses a normalized Gaussian random seed to generate the initial distribution in the generator to produce the new faceportraits. Table \ref{DCGANVal:Table} summarize the mean and standard deviation for all the faceportraits, which behavior is similar for all the generated faceportraits.
	
	\begin{table}[]
		\centering
		\caption{SSIM index normal distribution for the DCGAN generated faceportraits}
		\label{DCGANVal:Table}
		\renewcommand{\arraystretch}{0.6}
		\setlength{\tabcolsep}{2pt}
		\begin{adjustbox}{width=12cm,center}
			\begin{tabular}{|c|c|c|c|c|c|c|c|}
				\hline
				Faceportrait            & Statistic & Ball  & \begin{tabular}[c]{@{}c@{}}Inner\\ Race\end{tabular} & \begin{tabular}[c]{@{}c@{}}Load\\ Center\end{tabular} & \begin{tabular}[c]{@{}c@{}}Load\\ Opposite\end{tabular} & \begin{tabular}[c]{@{}c@{}}Load\\ Orthogonal\end{tabular} & Nominal \\ \hline
				\multirow{3}{*}{CWT}    & mean      & 0.961 & 0.946                                                & 0.969                                                 & 0.955                                                   & 0.971                                                     & 0.97    \\ \cline{2-8} 
				& std       & 0.003 & 0.005                                                & 0.008                                                 & 0.003                                                   & 0.002                                                     & 0.003   \\ \cline{2-8} 
				& Range     & 0.054 & 0.06                                                 & 0.052                                                 & 0.023                                                   & 0.011                                                     & 0.021   \\ \hline
				\multirow{3}{*}{CMR}    & mean      & 0.945 & 0.91                                                 & 0.76                                                  & 0.864                                                   & 0.959                                                     & 0.981   \\ \cline{2-8} 
				& std       & 0.003 & 0.006                                                & 0.168                                                 & 0.010                                                   & 0.003                                                     & 0.001   \\ \cline{2-8} 
				& Range     & 0.023 & 0.038                                                & 0.752                                                 & 0.071                                                   & 0.016                                                     & 0.008   \\ \hline
				\multirow{3}{*}{Gram}   & mean      & 0.027 & 0.892                                                & 0.721                                                 & 0.865                                                   & 0.939                                                     & 0.936   \\ \cline{2-8} 
				& std       & 0.004 & 0.006                                                & 0.099                                                 & 0.006                                                   & 0.004                                                     & 0.005   \\ \cline{2-8} 
				& Range     & 0.024 & 0.04                                                 & 0.464                                                 & 0.038                                                   & 0.024                                                     & 0.041   \\ \hline
				\multirow{3}{*}{Hankel} & mean      & 0.891 & 0.874                                                & 0.69                                                  & 0.838                                                   & 0.932                                                     & 0.947   \\ \cline{2-8} 
				& std       & 0.006 & 0.006                                                & 0.186                                                 & 0.012                                                   & 0.004                                                     & 0.004   \\ \cline{2-8} 
				& Range     & 0.04  & 0.039                                                & 0.798                                                 & 0.074                                                   & 0.03                                                      & 0.03    \\ \hline
				\multirow{3}{*}{Toep}   & mean      & 0.887 & 0.734                                                & 0.747                                                 & 0.833                                                   & 0.955                                                     & 0.97    \\ \cline{2-8} 
				& std       & 0.008 & 0.014                                                & 0.199                                                 & 0.012                                                   & 0.003                                                     & 0.003   \\ \cline{2-8} 
				& Range     & 0.064 & 0.082                                                & 0.875                                                 & 0.083                                                   & 0.021                                                     & 0.017   \\ \hline
				\multirow{3}{*}{Harr}   & mean      & 0.948 & 0.829                                                & 0.932                                                 & 0.848                                                   & 0.969                                                     & 0.968   \\ \cline{2-8} 
				& std       & 0.006 & 0.015                                                & 0.007                                                 & 0.021                                                   & 0.003                                                     & 0.003   \\ \cline{2-8} 
				& Range     & 0.056 & 0.094                                                & 0.0490                                                & 0.149                                                   & 0.018                                                     & 0.027   \\ \hline
			\end{tabular}
		\end{adjustbox}
	\end{table}

	%
	
	\subsection{CNN training for fault classification}
	A CNN network is trained to perform the failure detection between nominal and failure behaviors using the faceportraits balanced dataset generated by the DCGAN. The CNN is implemented in Matlab using the deep learning toolbox and is composed by three convolutional layers, two pooling stages with a ReLu activation function for the hidden layers and a sigmoid function in the output layer for the failure classification. One hundred epochs train the CNN with a learning rate of 0.001. The last layer has six outputs corresponding to the nominal case and the five failure behaviors inner race, ball, and outer race with center, opposite and orthogonal load. From the 12000 synthetic datasets, 3600 images were employed for the training process, using 300 images for each nominal and failure cases. The validation process employs 8400 images or 700 for each case.  After that, a second validation process is performed using the original dataset confirmed by 114 FacePortraits to verify the effectiveness of the CNN network after being trained with the balanced dataset. 
	
	\subsection{Faultface obtained results}
	The results of the faultFace methodology are summarized using the confusion matrix. It allows identifying the amount of true and false classifications considering if the classifier is confusing classes in the process. It is defined in terms of the true positives (TP), false positives (FP), false negatives (FN), and true negatives (TN) resulting from the fault detection algorithm. Table \ref{confMatrixCWT} and Table \ref{confMatHarr} present the confusion matrices obtained after applying the FaultFace methodology for each faceportrait. As can be observed, the CWT, CMR, Gram, Hankel, and Toeplitz FacePortraits gives a 100$\%$ matching for the validation data, indicating an excellent failure detection performance of the FaultFace methodology. However, in the case of the Haar FacePortrait, the obtained result shows that only the nominal, ball and load orthogonal behaviors have been detected correctly, while the inner race, load center, and load opposite failures are not well detected.
	\begin{table}
		\centering
		\caption{Confusion matrix for the FaultFace methodology with CWT, CMR, Gram, Hankel, and Toeplitz  FacePortraits}
		\label{confMatrixCWT}
		\renewcommand{\arraystretch}{0.6}
		\setlength{\tabcolsep}{2pt}
		\begin{adjustbox}{width=10cm,center}
			\begin{tabular}{c|c|c|c|c|c|c|c|}
				\cline{2-8}
				& \multicolumn{7}{c|}{\textbf{Target class}}                                                                                                                                                                                                                                                                                                                                     \\ \hline
				\multicolumn{1}{|c|}{\multirow{7}{*}{\rotatebox[origin=c]{90}{\textbf{Output class \phantom{cccc}}}}} & \textbf{}                                                          & \textbf{Ball} & \textbf{\begin{tabular}[c]{@{}c@{}}Inner\\ Race\end{tabular}} & \textbf{\begin{tabular}[c]{@{}c@{}}Load\\ Center\end{tabular}} & \textbf{\begin{tabular}[c]{@{}c@{}}Load\\ Opposite\end{tabular}} & \textbf{\begin{tabular}[c]{@{}c@{}}Load\\ Orthogonal\end{tabular}} & \textbf{Nominal} \\ \cline{2-8} 
				\multicolumn{1}{|c|}{}                                       & \textbf{Ball}                                                      & 28            & 0                                                             & 0                                                              & 0                                                                & 0                                                                  & 0                \\ \cline{2-8} 
				\multicolumn{1}{|c|}{}                                       & \textbf{\begin{tabular}[c]{@{}c@{}}Inner\\ Race\end{tabular}}      & 0             & 28                                                            & 0                                                              & 0                                                                & 0                                                                  & 0                \\ \cline{2-8} 
				\multicolumn{1}{|c|}{}                                       & \textbf{\begin{tabular}[c]{@{}c@{}}Load\\ Center\end{tabular}}     & 0             & 0                                                             & 23                                                             & 0                                                                & 0                                                                  & 0                \\ \cline{2-8} 
				\multicolumn{1}{|c|}{}                                       & \textbf{\begin{tabular}[c]{@{}c@{}}Load\\ Opposite\end{tabular}}   & 0             & 0                                                             & 0                                                              & 15                                                               & 0                                                                  & 0                \\ \cline{2-8} 
				\multicolumn{1}{|c|}{}                                       & \textbf{\begin{tabular}[c]{@{}c@{}}Load\\ Orthogonal\end{tabular}} & 0             & 0                                                             & 0                                                              & 0                                                                & 16                                                                 & 0                \\ \cline{2-8} 
				\multicolumn{1}{|c|}{}                                       & \textbf{Nominal}                                                   & 0             & 0                                                             & 0                                                              & 0                                                                & 0                                                                  & 4                \\ \hline
			\end{tabular}
		\end{adjustbox}
	\end{table}
	\begin{table}
		\centering
		\caption{Confusion matrix for the FaultFace methodology with Haar FacePortraits}
		\label{confMatHarr}
		\renewcommand{\arraystretch}{0.6}
		\setlength{\tabcolsep}{2pt}
		\begin{adjustbox}{width=10cm,center}
			\begin{tabular}{c|c|c|c|c|c|c|c|}
				\cline{2-8}
				& \multicolumn{7}{c|}{\textbf{Target class}}                                                                                                                                                                                                                                                                                                                                     \\ \hline
				\multicolumn{1}{|c|}{\multirow{7}{*}{\rotatebox[origin=c]{90}{\textbf{Output class \phantom{cccc} }}}} & \textbf{}                                                          & \textbf{Ball} & \textbf{\begin{tabular}[c]{@{}c@{}}Inner\\ Race\end{tabular}} & \textbf{\begin{tabular}[c]{@{}c@{}}Load\\ Center\end{tabular}} & \textbf{\begin{tabular}[c]{@{}c@{}}Load\\ Opposite\end{tabular}} & \textbf{\begin{tabular}[c]{@{}c@{}}Load\\ Orthogonal\end{tabular}} & \textbf{Nominal} \\ \cline{2-8} 
				\multicolumn{1}{|c|}{}                                       & \textbf{Ball}                                                      & 28            & 0                                                             & 0                                                              & 0                                                                & 0                                                                  & 0                \\ \cline{2-8} 
				\multicolumn{1}{|c|}{}                                       & \textbf{\begin{tabular}[c]{@{}c@{}}Inner\\ Race\end{tabular}}      & 0             & 0                                                            & 0                                                              & 0                                                                & 28                                                                  & 0                \\ \cline{2-8} 
				\multicolumn{1}{|c|}{}                                       & \textbf{\begin{tabular}[c]{@{}c@{}}Load\\ Center\end{tabular}}     & 0             & 0                                                             & 0                                                             & 0                                                                & 23                                                                  & 0                \\ \cline{2-8} 
				\multicolumn{1}{|c|}{}                                       & \textbf{\begin{tabular}[c]{@{}c@{}}Load\\ Opposite\end{tabular}}   & 0             & 0                                                             & 0                                                              & 0                                                               & 15                                                                  & 0                \\ \cline{2-8} 
				\multicolumn{1}{|c|}{}                                       & \textbf{\begin{tabular}[c]{@{}c@{}}Load\\ Orthogonal\end{tabular}} & 0             & 0                                                             & 0                                                              & 0                                                                & 16                                                                 & 0                \\ \cline{2-8} 
				\multicolumn{1}{|c|}{}                                       & \textbf{Nominal}                                                   & 0             & 0                                                             & 0                                                              & 0                                                                & 0                                                                  & 4                \\ \hline
			\end{tabular}
		\end{adjustbox}
	\end{table}
	
	\subsection{Results analysis of the faultFace methodology using DCGAN networks}
	The performance of the faultFace Methodology is quantified using the confusion matrix. Three indices given by \eqref{confMatMetrics} are calculated, the accuracy A, which establishes the fault rate of the method, the coverage C, which indicates the overall effectiveness of the classifier, and the harmonic mean F, which defines the deviation of the data from the mean. 
	\begin{eqnarray}
	\label{confMatMetrics}
	A=\frac{TP}{TP+FP} \quad C=\frac{TP}{TP+FN} \quad F=\frac{2AC}{A+C}_.
	\end{eqnarray}
	The proposed performance indices are summarized in Table.\ref{DCGANConfPerf}. As can be observed, for the CWT and CMR FacePortraits, the FaultFace methodology gives an accuracy, coverage, and harmonic mean of 1. It means that the synthetic dataset created using the DCGAN has excellent performance for training the CNN for failure classification combined with a good generalization from the CNN. On the other hand, the performance indices show that the accuracy and consistency of the FaultFace method change when the Haar FacePortrait is employed. It can be noticed in the fact that only the nominal and ball failure has been correctly classified, but in the case of Inner race, and outer race with the center, opposite and orthogonal load the algorithm cannot differentiate between the failures.
	\begin{table} [h]
		\centering
		\caption{Performance metrics for the  FaultFace methodology for each face portraits}
		\label{DCGANConfPerf}
		\renewcommand{\arraystretch}{0.62}
		\setlength{\tabcolsep}{2pt}
		\begin{tabular}{|c|c|c|c|c|}
			\hline
			\multirow{2}{*}{\textbf{\begin{tabular}[c]{@{}c@{}}Face\\ Portrait\end{tabular}}} & \multirow{2}{*}{\textbf{Failure}} & \multicolumn{3}{c|}{\textbf{Index}}                                                                       \\ \cline{3-5} 
			&                                   & \textbf{Accuracy} & \textbf{Coverage} & \textbf{\begin{tabular}[c]{@{}c@{}}Harmonic\\  mean\end{tabular}} \\ \hline
			\multirow{6}{*}{\textbf{\begin{tabular}[c]{@{}c@{}}CWT\\CMR\\Gram\\Hankel\\Toep\end{tabular}}}                                                     & Nominal                           & 1                 & 1                 & 1                                                                 \\ \cline{2-5} 
			& Ball                              & 1                 & 1                 & 1                                                                 \\ \cline{2-5} 
			& Inner Race                        & 1                 & 1                 & 1                                                                 \\ \cline{2-5} 
			& Load Center                       & 1                 & 1                 & 1                                                                 \\ \cline{2-5} 
			& Load Opposite                     & 1                 & 1                 & 1                                                                 \\ \cline{2-5} 
			& Load Orthogonal                   & 1                 & 1                 & 1                                                                 \\ \hline
			\multirow{6}{*}{\textbf{Haar}}                                                    & Nominal                           &  1                 & 1                  &1                                                                 \\ \cline{2-5} 
			& Ball                              &  1                 &1                   &1                                                                   \\ \cline{2-5} 
			& Inner Race                        & 0                  &0                   & 0                                                                   \\ \cline{2-5} 
			& Load Center                       & 0                  &0                   &  0                                                                 \\ \cline{2-5} 
			& Load Opposite                     & 0                  &0                   & 0                                                                  \\ \cline{2-5} 
			& Load Orthogonal                   &1                  &0.238                   &0.379                                                                   \\ \hline
			
		\end{tabular}
	\end{table}
	For the orthogonal load, the accuracy is one because the algorithm can recognize all the samples related to this behavior; however, the consistency is close to 0.238 because the classification algorithm confuses these with the orthogonal case. Likewise, the harmonic mean of 0.379 indicates a high data dispersion of the fault detector using this face portrait. A  possible cause for this behavior is that the Haar wavelet does not represent adequately in the time-frequency domain the different features of that failure behaviors. For this reason, it is possible to say that the choice of the face portrait is not a trivial task and has a significant effect over the fault detection final performance.
	
	\section{FaultFace methodology using GAN network}
\begin{figure}
		\centering
		\includegraphics[width=1\textwidth,height=0.25\textheight]{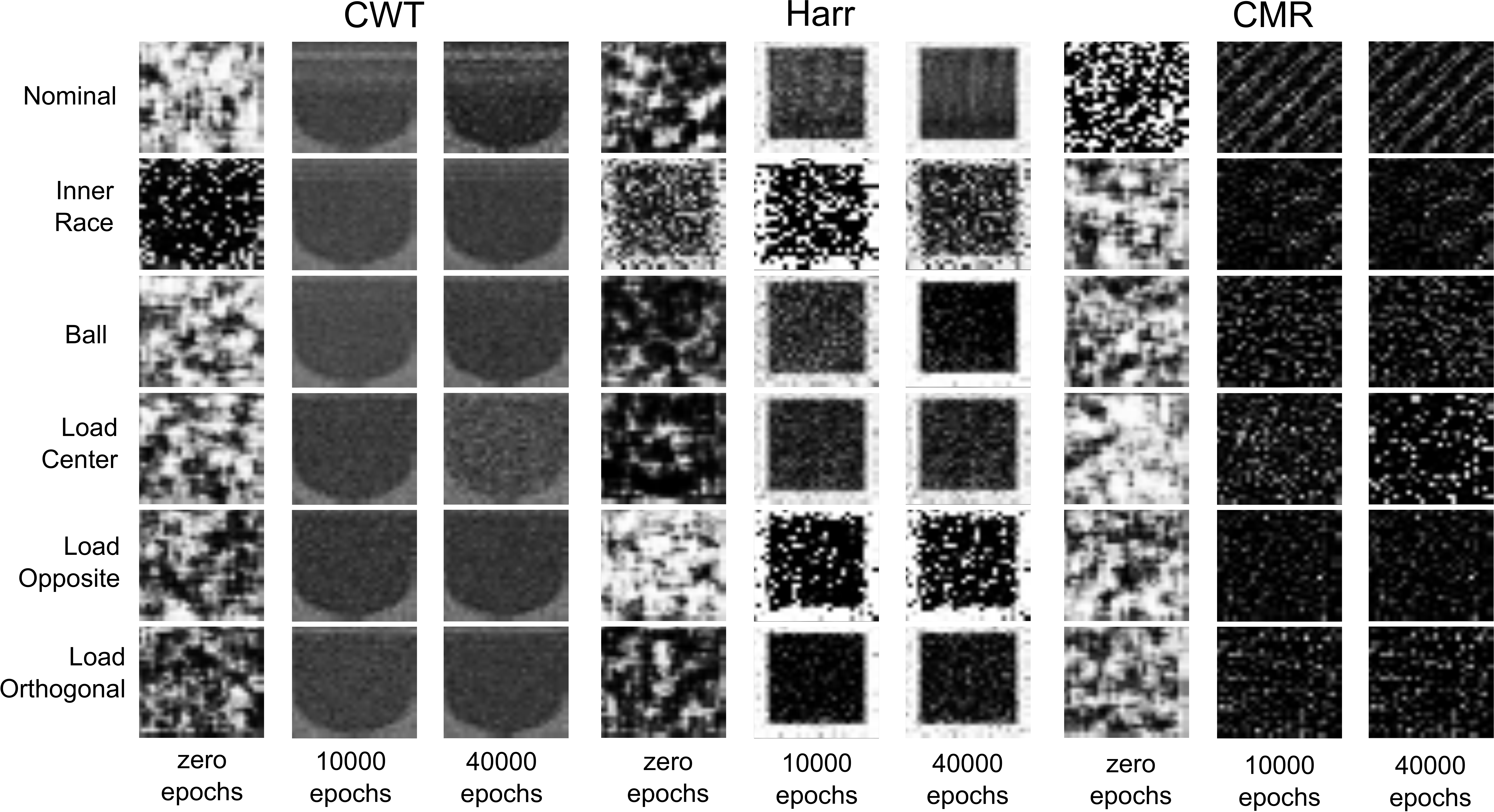}	
		\caption[caption]{GAN generated CWT, Harr, and CMR face portraits at zero, 10000, and 40000 epochs}
		\label{fpGAN1}
	\end{figure}

\begin{figure}
		\centering
		\includegraphics[width=1\textwidth,height=0.25\textheight]{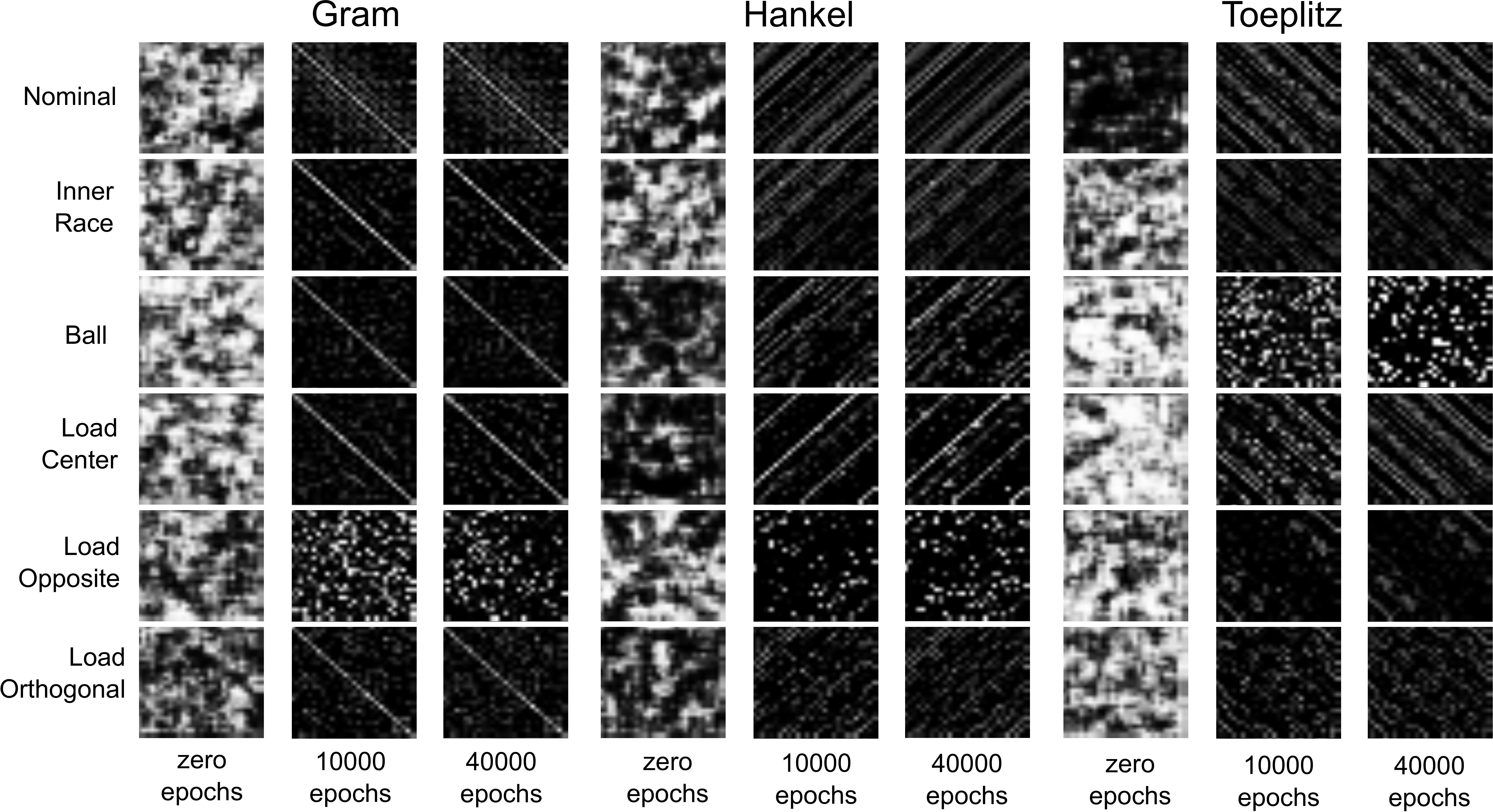}	
		\caption[caption]{GAN generated Gram, Hankel, and Toeplitz face portraits at zero, 10000, and 40000 epochs }
		\label{fpGAN2}
	\end{figure}

	The FaultFace methodology is performed using a GAN network in order to compare with obtained results using the DCGAN for dataset balancing tasks. In this case, the GAN network employs multilayer perceptron networks for the discriminator and the generator. The structure of the generator uses three full connected layers of 256, 512, and 1024 neurons respectively and an output layer of 784 neurons to fit with the 28x28 generated faceportrait dimensions. The initial minibatch input size is 100 samples generated using Gaussian distribution. The activation function for the first two layers uses LeakyRelu as activation function, and hyperbolic tangent for the output layer. A batch normalization operator is included at the output of each activation function. The discriminator network is conformed by two fully connected layers with 512 and 256 neurons with leakyRelu activation function, and an output layer with sigmoid activation function to decide between a fake and correct image. A GAN network is trained for the nominal an failure behaviors of the ball-bearing system, for 40000 iterations using adam optimizer with a learning rate of 0.0002 with decay rate of 0.5. The obtained faceportraits obtained using GAN networks are shown in Fig.\ref{fpGAN1} and Fig.\ref{fpGAN2}.
	\par
	The quality of the new faceportraits generated with the GAN network is measured with the SSIM index presented in section 4.4. Fig.\ref{dcganVal} shows a boxplot of the SSIM index calculated for the nominal CWT and Hankel faceportraits for the nominal and fault behaviors and Table.\ref{GANVal:Table} summaries all the results obtained for the GAN network. It can be observed that the GAN network SSIM index has a big dispersion on the balanced dataset for all the cases, indicating that the generated data from the GAN diverges considerably from the original data, which will have an effect on the fault detection task.
	
	\begin{figure}
		\centering
		\subfloat[]{\includegraphics[width=0.5\textwidth,height=0.2\textheight]{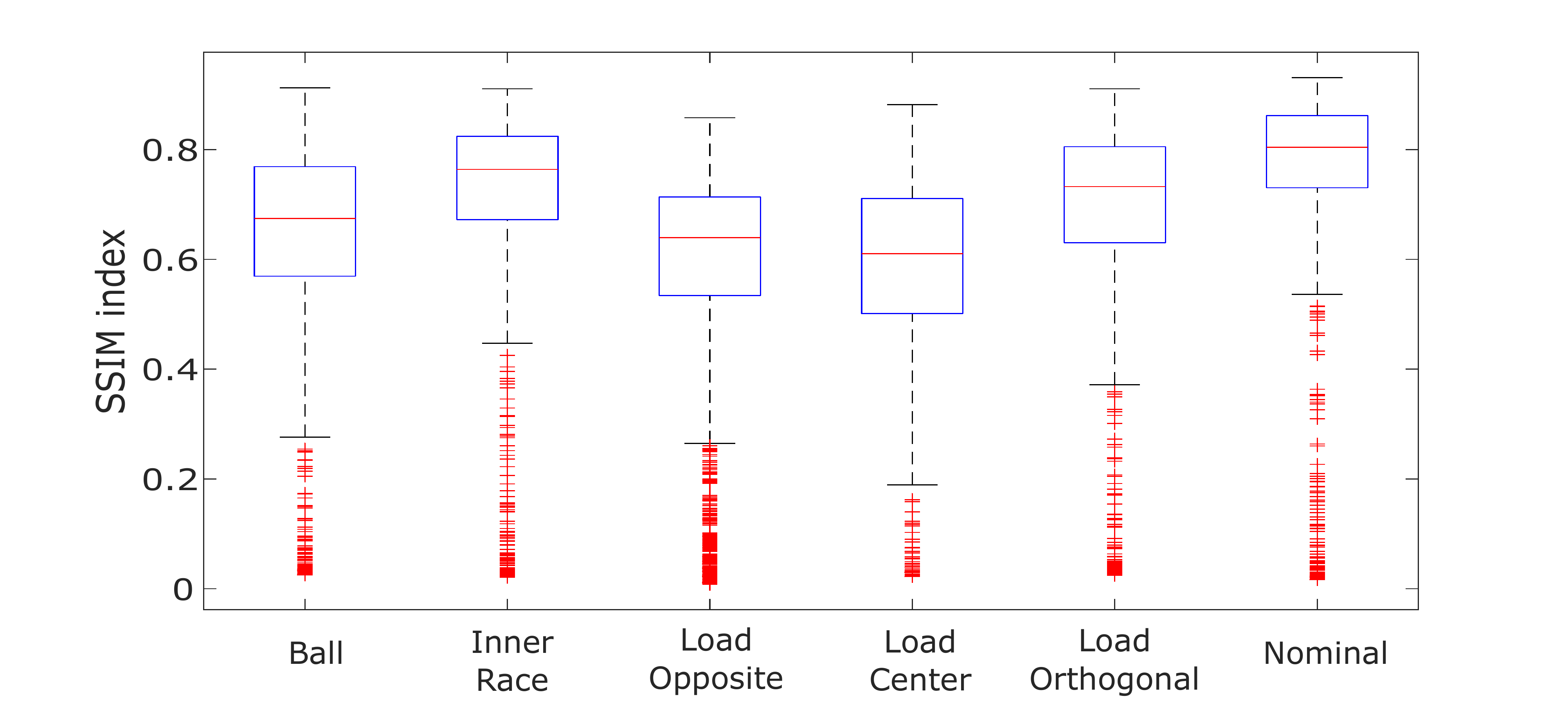}}
		\subfloat[]{\includegraphics[width=0.5\textwidth,height=0.2\textheight]{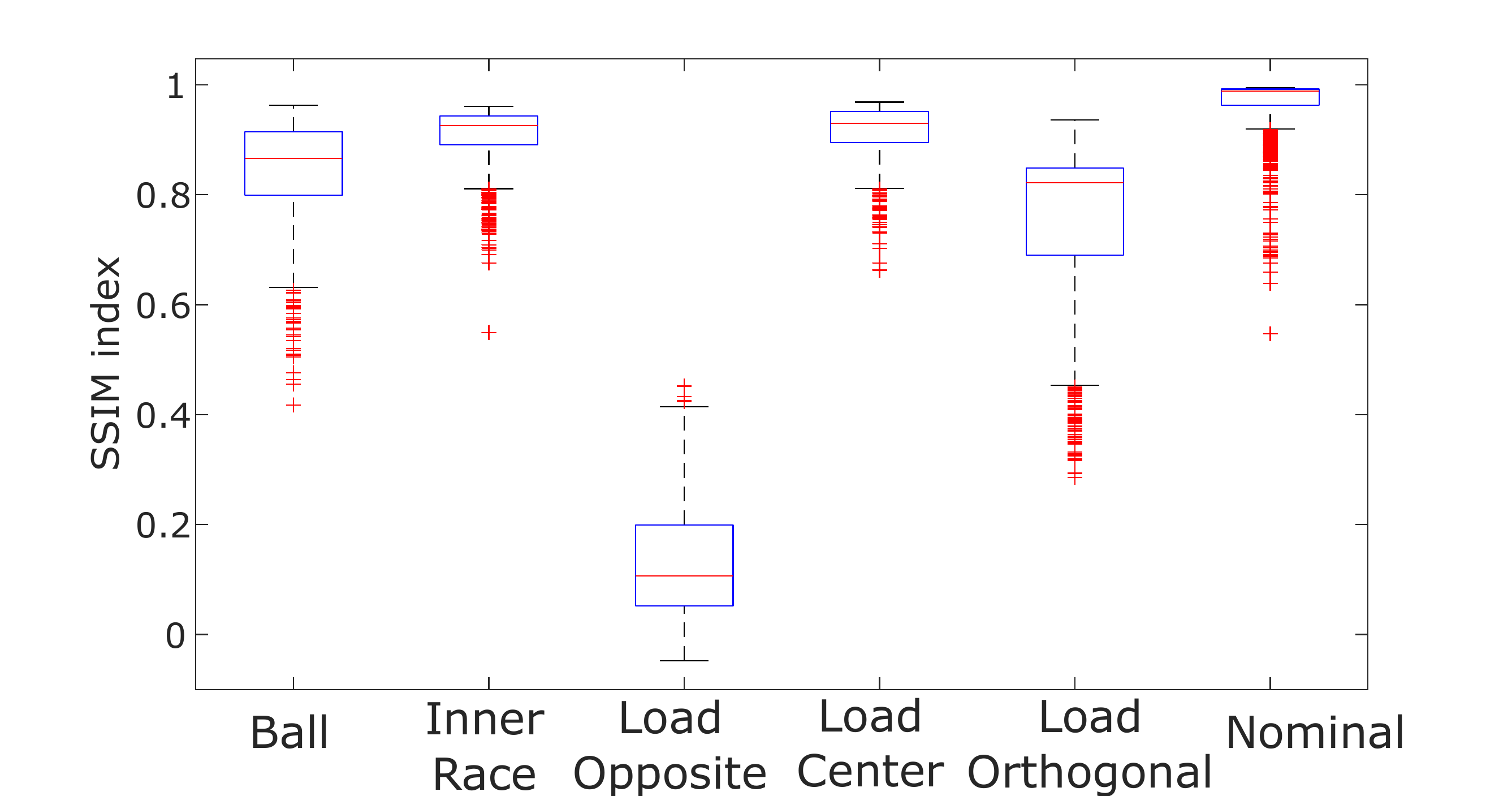}}
		\caption[caption]{SSIM index for quality assesment of the balanced dataset produced by the DCGAN network for the a) CWT and b)Hankel faceportraits}
		\label{ganVal}
	\end{figure}
	
	\begin{table}[]
		\centering
		\caption{SSIM index normal distribution for the GAN generated faceportraits}
		\label{GANVal:Table}
		\renewcommand{\arraystretch}{0.6}
		\setlength{\tabcolsep}{2pt}
		\begin{adjustbox}{width=12cm,center}
			\begin{tabular}{|c|c|c|c|c|c|c|c|}
				\hline
				Faceportrait            & Statistic & Ball  & \begin{tabular}[c]{@{}c@{}}Inner\\ Race\end{tabular} & \begin{tabular}[c]{@{}c@{}}Load\\ Center\end{tabular} & \begin{tabular}[c]{@{}c@{}}Load\\ Opposite\end{tabular} & \begin{tabular}[c]{@{}c@{}}Load\\ Orthogonal\end{tabular} & Nominal \\ \hline
				\multirow{3}{*}{CWT}    & mean      & 0.625 & 0.663                                                & 0.561                                                 & 0.582                                                   & 0.673                                                     & 0.715   \\ \cline{2-8} 
				& std       & 0.216 & 0.261                                                & 0.236                                                 & 0.188                                                   & 0.211                                                     & 0.257   \\ \cline{2-8} 
				& Range     & 0.888 & 0.89                                                 & 0.85                                                  & 0.86                                                    & 0.887                                                     & 0.915   \\ \hline
				\multirow{3}{*}{CMR}    & mean      & 0.85  & 0.914                                                & 0.793                                                 & 0.512                                                   & 0.942                                                     & 0.887   \\ \cline{2-8} 
				& std       & 0.108 & 0.06                                                 & 0.206                                                 & 0.221                                                   & 0.040                                                     & 0.228   \\ \cline{2-8} 
				& Range     & 0.492 & 0.44                                                 & 0.924                                                 & 0.832                                                   & 0.256                                                     & 0.968   \\ \hline
				\multirow{3}{*}{Gram}   & mean      & 0.859 & 0.77                                                 & 0.090                                                 & 0.867                                                   & 0.855                                                     & 0.91    \\ \cline{2-8} 
				& std       & 0.072 & 0.119                                                & 0.042                                                 & 0.07                                                    & 0.086                                                     & 0.069   \\ \cline{2-8} 
				& Range     & 0.461 & 0.644                                                & 0.208                                                 & 0.389                                                   & 0.511                                                     & 0.449   \\ \hline
				\multirow{3}{*}{Hankel} & mean      & 0.844 & 0.907                                                & 0.132                                                 & 0.917                                                   & 0.752                                                     & 0.963   \\ \cline{2-8} 
				& std       & 0.094 & 0.052                                                & 0.102                                                 & 0.047                                                   & 0.145                                                     & 0.057   \\ \cline{2-8} 
				& Range     & 0.545 & 0.412                                                & 0.50                                                  & 0.306                                                   & 0.65                                                      & 0.448   \\ \hline
				\multirow{3}{*}{Toep}   & mean      & 0.07  & 0.7                                                  & 0.671                                                 & 0.162                                                   & 0.719                                                     & 0.887   \\ \cline{2-8} 
				& std       & 0.037 & 0.128                                                & 0.225                                                 & 0.021                                                   & 0.126                                                     & 0.079   \\ \cline{2-8} 
				& Range     & 0.29  & 0.642                                                & 0.905                                                 & 0.108                                                   & 0.6                                                       & 0.496   \\ \hline
				\multirow{3}{*}{Harr}   & mean      & 0.759 & 0.417                                                & 0.328                                                 & 0.745                                                   & 0.733                                                     & 0.781   \\ \cline{2-8} 
				& std       & 0.055 & 0.116                                                & 0.075                                                 & 0.037                                                   & 0.036                                                     & 0.207   \\ \cline{2-8} 
				& Range     & 0.533 & 0.536                                                & 0.432                                                 & 0.293                                                   & 0.245                                                     & 0.883   \\ \hline
			\end{tabular}
		\end{adjustbox}
	\end{table}
	
\begin{figure}
		\centering
		\includegraphics[width=0.8\textwidth,height=0.3\textheight]{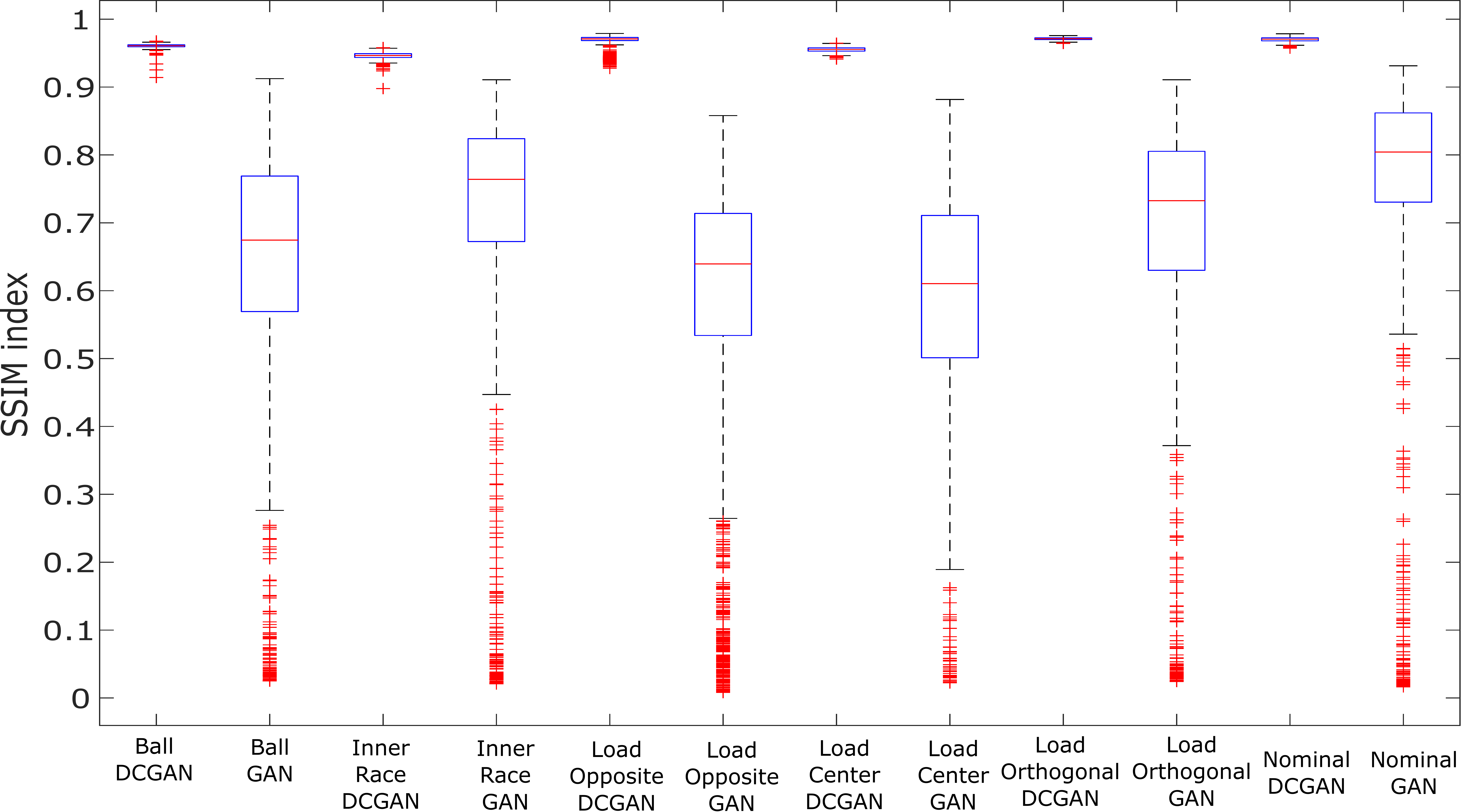}
		\caption[caption]{SSIM index for quality assessment of the balanced datasets produced by the DCGAN and GAN networks for the CWT faceportrait}
		\label{GANvsDCGAN}
	\end{figure}

	In addition, a comparison between the SSIM of GAN and DCGAN generated balanced dataset is presented on Fig.\ref{GANvsDCGAN} for the CWT faceportrait. It can be observed that DCGAN network produces more accurate new data from the original dataset compared with the GAN network, with higher mean SSIM value and less dispersion of the data distribution.
	
	Table.\ref{GANConfPerf} shows the accuracy, precision, and harmonic mean from the confusion matrices obtained each faceportrait using the FaultFace methodology with the GAN network balanced dataset. As can be observed, only the CMR faceportrait returns a 100\% accuracy on the failure classification task.  For the CWT and Toep faceportraits, the CNN makes an incorrect differentiation of the load center failure, confusing it with load opposite and inner race faults respectively. In the case of Gram and Hankel faceportraits, the classifier does not recognize properly the load opposite fault. Finally, the Harr faceportrait has similar classification problems as result with the DCGAN network due to the absence of features offered by this faceportrait for the fault detection task. Thus, it is possible to say that the DCGAN network is a good option for dataset balancing compared with GAN network for fault detection applications.
	
	\begin{table}
		\centering
		\caption{Performance metrics for the  FaultFace methodology for each face portraits generated using GAN network}
		\label{GANConfPerf}
		\renewcommand{\arraystretch}{0.62}
		\setlength{\tabcolsep}{2pt}
		\begin{adjustbox}{width=9cm,center}
			\begin{tabular}{|c|c|c|c|c|}
				\hline
				\multirow{2}{*}{\textbf{\begin{tabular}[c]{@{}c@{}}Face\\ Portrait\end{tabular}}} & \multirow{2}{*}{\textbf{Failure}} & \multicolumn{3}{c|}{\textbf{Index}}                                                                       \\ \cline{3-5} 
				&                                   & \textbf{Accuracy} & \textbf{Coverage} & \textbf{\begin{tabular}[c]{@{}c@{}}Harmonic\\  mean\end{tabular}} \\ \hline
				\multirow{6}{*}{\textbf{\begin{tabular}[c]{@{}c@{}}CMR\end{tabular}}}                                                     & Nominal                           & 1                 & 1                 & 1                                                                 \\ \cline{2-5} 
				& Ball                              & 1                 & 1                 & 1                                                                 \\ \cline{2-5} 
				& Inner Race                        & 1                 & 1                 & 1                                                                 \\ \cline{2-5} 
				& Load Center                       & 1                 & 1                 & 1                                                                 \\ \cline{2-5} 
				& Load Opposite                     & 1                 & 1                 & 1                                                                 \\ \cline{2-5} 
				& Load Orthogonal                   & 1                 & 1                 & 1                                                                 \\ \hline
				\multirow{6}{*}{\textbf{CWT}}                                                    & Nominal                           &  1                 & 1                  &1                                                                 \\ \cline{2-5} 
				& Ball                              & 1                 &1                   &1                                                                  \\ \cline{2-5} 
				& Inner Race                        & 1                  &1                   & 0                                                                   \\ \cline{2-5} 
				& Load Center                       &0                  &0                   &  0                                                                 \\ \cline{2-5} 
				& Load Opposite                     &1                 &0.395                   & 0.566                                                                \\ \cline{2-5} 
				& Load Orthogonal                   &1                  &1                  &1                                                                  \\ \hline
				
				\multirow{6}{*}{\textbf{Gram}}                                                    & Nominal                           &  1                 & 1                  &1                                                                 \\ \cline{2-5} 
				& Ball                              &  0.6585                 &1                   &0.7904                                                                   \\ \cline{2-5} 
				& Inner Race                        & 1                  &1                  & 1                                                                  \\ \cline{2-5} 
				& Load Center                       & 1                  &1                   &  1                                                                 \\ \cline{2-5} 
				& Load Opposite                     & 0                  &0                   & 0                                                                  \\ \cline{2-5} 
				& Load Orthogonal                   &1                  &1                   &1                                                                  \\ \hline
				
				\multirow{6}{*}{\textbf{Hankel}}                                                    & Nominal                           &  1                 & 1                  &1                                                                 \\ \cline{2-5} 
				& Ball                              &  1                 &1                   &1                                                                   \\ \cline{2-5} 
				& Inner Race                        & 0.6511                  &1                   & 0.7878                                                                   \\ \cline{2-5} 
				& Load Center                       & 1                  &1                   &  1                                                                 \\ \cline{2-5} 
				& Load Opposite                     & 0                  &0                   & 0                                                                  \\ \cline{2-5} 
				& Load Orthogonal                   &1                  &1                  &1                                                                   \\ \hline
				
				\multirow{6}{*}{\textbf{Harr}}                                                    & Nominal                           &  1                 & 1                  &1                                                                 \\ \cline{2-5} 
				& Ball                              &  0.4757                 &1                   &0.6447                                                                   \\ \cline{2-5} 
				& Inner Race                        & 0                  &0                   & 0                                                                   \\ \cline{2-5} 
				& Load Center                       & 1                  &1                   &  1                                                                 \\ \cline{2-5} 
				& Load Opposite                     & 0                  &0                   & 0                                                                  \\ \cline{2-5} 
				& Load Orthogonal                   &0                  &0                   &0                                                                   \\ \hline
				
				\multirow{6}{*}{\textbf{Toep}}                                                    & Nominal                           &  1                 & 1                  &1                                                                 \\ \cline{2-5} 
				& Ball                              &  0                 &0                   &0                                                                   \\ \cline{2-5} 
				& Inner Race                        & 0.5490                  &0.5                   &0.5233                                                                   \\ \cline{2-5} 
				& Load Center                       & 0                  &0                   &  0                                                                 \\ \cline{2-5} 
				& Load Opposite                     & 1                  &1                   & 1                                                                  \\ \cline{2-5} 
				& Load Orthogonal                   &1                  &1                   &1                                                                   \\ \hline

			\end{tabular}
		\end{adjustbox}
	\end{table}

	\section{FaultFace comparison with other methodologies}
	
	An LSTM and a SVM with autoencoder networks are designed to perform the fault detection task for the ball-bearing system and compare its performance with the faultFace methodology. Likewise, the FaultFace methodology is also compared with results reported on the literature for the same ball-bearing benchmark system \cite{b26}.
	
	\subsection{LSTM Network}
	
	A Long Short-Term Memory network (LSTM) \cite{lstmOriginal} is used for the vibration time-series classification. The LSTM architecture is composed by a unidirectional LSTM layer of 100 hidden units, with a input size of 1000 samples, combined a fully connected layer with softmax activation function. The output layer has six outputs for the nominal and the five fault behaviors. The LSTM is trained for 50 epochs, with a minibatch size of 100 samples. The training and validation datasets are composed by 12 and 102 timseries respectively divided in minibatches with variable length between 80 and 100 values. The confusion matrix metrics for the LSTM network are presented in Table\ref{confMatLSTM}. It can be observed that using LSTM for the ball bearing fault detection problem, an overall accuracy of 69\% is reached. Also, the LSTM network exhibit some challenges classifying between the load disturbances cases (center, orthogonal, opposite).
	
	\begin{table}[h]
		\centering
		\caption{Performance metrics for the  FaultFace methodology for each face portraits}
		\label{confMatLSTM}
		\renewcommand{\arraystretch}{0.62}
		\setlength{\tabcolsep}{2pt}
		\begin{adjustbox}{width=9cm,center}
			\begin{tabular}{|c|c|c|c|c|}
				\hline
				\multirow{2}{*}{\textbf{\begin{tabular}[c]{@{}c@{}}Technique\end{tabular}}} & \multirow{2}{*}{\textbf{Failure}} & \multicolumn{3}{c|}{\textbf{Index}}                                                                       \\ \cline{3-5} 
				&                                   & \textbf{Accuracy} & \textbf{Coverage} & \textbf{\begin{tabular}[c]{@{}c@{}}Harmonic\\  mean\end{tabular}} \\ \hline
				\multirow{6}{*}{\textbf{\begin{tabular}[c]{@{}c@{}}LSTM\end{tabular}}}                                                     & Nominal                           & 0.5                 & 1                 & 0.6667                                                                 \\ \cline{2-5} 
				& Ball                              & 1                 & 1                 & 1                                                                 \\ \cline{2-5} 
				& Inner Race                        &1                 & 1                 & 1                                                                 \\ \cline{2-5} 
				& Load Center                       & 0.65                 & 1                 & 0.7878                                                                 \\ \cline{2-5} 
				& Load Opposite                     &0.5                 & 1                 & 0.6667                                                                 \\ \cline{2-5} 
				& Load Orthogonal                   & 0                 & 0                 & 0                                                                 \\ \hline
			\end{tabular}
		\end{adjustbox}
	\end{table}
	
	\subsection{SVM with Autoencoder}
	An autoencoder with a support vector machine (SVM) is implemented for the fault detection of the ball bearing system. The Autoencoder reduces the faceportraits dimensionality using a hidden layer of 100 neurons and an output layer of 10 output features. It is trained for 1000 epochs with L2 weight regularization of 0.004. After that, the SVM is trained using the output of the Autoencoder to perform the fault detection task. Table.\ref{SVMConfPerf} shows the accuracy, coverage, and harmonic mean F metrics calculated for the ball bearing system using the unbalanced dataset, and the balanced datasets using the GAN and DCGAN networks. The model trained with the unbalanced dataset employs 57 faceportraits for training and 57 for validation, from a total of 114 faceportraits. Thus, an accuracy of about 70\% is reached. In the case of the balanced datasets generated with GAN and DCGAN networks, each dataset has 6000 faceportraits, which 3000 were used for training and 300 for validation. For the balanced dataset with the DCGAN network, the accuracy reached is almost 100\% for all the cases. In the case of the GAN network, the balanced dataset is about 85\%, improving the result obtained with the unbalanced dataset. So, the dataset balancing operation performed by the DCGAN and GAN network is essential to improve the fault detection task accuracy.
	\begin{table}[]
		\centering
		\caption{Performance metrics for the SVM with autoencoder}
		\renewcommand{\arraystretch}{0.62}
		\setlength{\tabcolsep}{2pt}	
		\begin{adjustbox}{width=12cm, height=2.2cm,center}
			\label{SVMConfPerf}
			\begin{tabular}{|c|c|c|c|c|c|c|c|c|c|}
				\hline
				\multirow{3}{*}{FacePortrait} & \multicolumn{9}{c|}{SVM with Autoencoder}                                                                                                                                                                                                                                            \\ \cline{2-10} 
				& \multicolumn{3}{c|}{\begin{tabular}[c]{@{}c@{}}Unbalanced \\ dataset train\end{tabular}} & \multicolumn{3}{c|}{\begin{tabular}[c]{@{}c@{}}Balanced GAN\\  dataset train\end{tabular}} & \multicolumn{3}{c|}{\begin{tabular}[c]{@{}c@{}}Balanced DCGAN\\  dataset train\end{tabular}} \\ \cline{2-10} 
				& Accuracy     & Coverage     & \begin{tabular}[c]{@{}c@{}}Harmonic\\ mean\end{tabular}    & Accuracy      & Coverage     & \begin{tabular}[c]{@{}c@{}}Harmonic\\ mean\end{tabular}     & Accuracy      & Coverage      & \begin{tabular}[c]{@{}c@{}}Harmonic\\ mean\end{tabular}      \\ \hline
				CMR                           & 0.772        & 1            & 0.8713                                                     & 0.927         & 0.9411       & 0.9340                                                      & 1             & 1             & 1                                                            \\ \hline
				CWT                           & 0.7189       & 1            & 0.8364                                                     & 0.87          & 0.834        & 0.85                                                        & 1             & 1             & 1                                                            \\ \hline
				Gram                          & 0.807        & 1            & 0.8931                                                     & 0.873         & 0.9360       & 0.9034                                                      & 1             & 1             & 1                                                            \\ \hline
				Hankel                        & 0.684        & 1            & 0.8123                                                     & 0.97          & 0.997        & 0.7179                                                      & 0.985         & 0.9879        & 0.983                                                        \\ \hline
				Harr                          & 0.684        & 1            & 0.8123                                                     & 0.794         & 0.96         & 0.8691                                                      & 0.99          & 1             & 1                                                            \\ \hline
				Toep                          & 0.86         & 1            & 0.9247                                                     & 0.856         & 0.95         & 0.9                                                         & 1             & 1             & 1                                                            \\ \hline
			\end{tabular}
		\end{adjustbox}
	\end{table}
	
	\subsection{Fault detection technique for ball-bearing in the literature}
	A review about another methodologies for the ball bearing fault detection on the benchmark system \cite{b26} was performed to made a comparison with the FaultFace method \cite{b19},\cite{b20},\cite{b31}-\cite{b33}. In Table \ref{BBComparison}, a summary of the different reviewed papers is presented, which employ supervised learning in many cases, some unsupervised and another one use traditional vibration methods like fast Fourier transform. In \cite{b34} is presented a supervised machine learning approach using SVM for failure detection with the best fitness of 99$\%$. Besides, \cite{b31} and \cite{b32} present the use of fractal theory for feature extraction and classification of failure with an accuracy of 98.4\% and 96.59\% respectively. On the other hand, \cite{b33} employs traditional Fourier analysis to detect the different failure behaviors based on the kurtosis of the frequency spectrum of the vibration signal. In the particular case of \cite{b19} and \cite{b20}, both techniques employ unsupervised learning combined with deep learning techniques for failure classification of the ball bearing system. On \cite{b19}, the Kmeans algorithm is combined with a GAN network and an autoencoder to create a dimensional reduction of the dataset to detect failures reaching a peak accuracy of 94.69\%. In \cite{b20}, a Deep Neural Network is employed for the fault detection, beginning with a feature extraction from the frequency spectrum of the signals and the use of Principal Component Analysis (PCA) to reduce the data dimension. After that, the network is trained based on the 3D PCA map of each signal. The accuracy achieved is 100\% for seven clusters. In addition, the LSTM and the SVM with autoencoder techniques proposed in this paper are included in the table with accuracy of 90\% and 69\% respectively.
	
	\begin{table}
		\centering
		\caption{Comparison between different failure detection methods for ball bearing elements}
		\renewcommand{\arraystretch}{0.62}
		\setlength{\tabcolsep}{2pt}
		\label{BBComparison}
		\begin{adjustbox}{width=9cm,center}
			\begin{tabular}{cllll}
				\hline
				\multicolumn{1}{|c|}{\textbf{Paper}} & \multicolumn{1}{c|}{\textbf{Type}} & \multicolumn{1}{c|}{\textbf{\begin{tabular}[c]{@{}c@{}}Classification \\ techniques\end{tabular}}}                 & \multicolumn{1}{c|}{\textbf{University}}                                                  & \multicolumn{1}{c|}{\textbf{\begin{tabular}[c]{@{}c@{}}Best \\ Accuracy\end{tabular}}} \\ \hline
				\multicolumn{1}{|c|}{FaultFace}      & \multicolumn{1}{c|}{Supervised}    & \multicolumn{1}{c|}{\begin{tabular}[c]{@{}c@{}}DCGAN with\\  CNN network\end{tabular}}                             & \multicolumn{1}{c|}{\begin{tabular}[c]{@{}c@{}}U. of California\\  Merced\end{tabular}}   & \multicolumn{1}{c|}{100\%}                                                             \\ \hline
				\multicolumn{1}{|c|}{{[}20{]}}       & \multicolumn{1}{c|}{Unsupervised}  & \multicolumn{1}{c|}{\begin{tabular}[c]{@{}c@{}}Deep neural \\ network\end{tabular}}                                & \multicolumn{1}{c|}{\begin{tabular}[c]{@{}c@{}}Tianjin \\ Polytechincal U\end{tabular}}   & \multicolumn{1}{c|}{100\%}                                                             \\ \hline
				\multicolumn{1}{|c|}{{[}34{]}}       & \multicolumn{1}{c|}{Supervised}    & \multicolumn{1}{c|}{\begin{tabular}[c]{@{}c@{}}Minimum entropy \\ deconvolution \\ with SVM\end{tabular}}          & \multicolumn{1}{c|}{\begin{tabular}[c]{@{}c@{}}U. of\\  Pardubice\end{tabular}}           & \multicolumn{1}{c|}{99.30\%}                                                           \\ \hline
				\multicolumn{1}{|c|}{{[}31{]}}       & \multicolumn{1}{c|}{Supervised}    & \multicolumn{1}{c|}{\begin{tabular}[c]{@{}c@{}}Fractal box \\ counting dimension\end{tabular}}                     & \multicolumn{1}{c|}{\begin{tabular}[c]{@{}c@{}}Harbin \\ Engineering U.\end{tabular}}     & \multicolumn{1}{c|}{98.40\%}                                                           \\ \hline
				\multicolumn{1}{|c|}{{[}32{]}}       & \multicolumn{1}{c|}{Supervised}    & \multicolumn{1}{c|}{\begin{tabular}[c]{@{}c@{}}Multifractal\\  and gray relation\end{tabular}}                     & \multicolumn{1}{c|}{\begin{tabular}[c]{@{}c@{}}Shanghai \\ Dianji U.\end{tabular}}        & \multicolumn{1}{c|}{96.59\%}                                                           \\ \hline
				\multicolumn{1}{|c|}{{[}19{]}}       & \multicolumn{1}{c|}{Unsupervised}  & \multicolumn{1}{c|}{\begin{tabular}[c]{@{}c@{}}Kmeans, with \\ Generative\\  adversarial autoencoder\end{tabular}} & \multicolumn{1}{c|}{\begin{tabular}[c]{@{}c@{}}Huazhong U. of \\ Technology\end{tabular}} & \multicolumn{1}{c|}{94.69\%}                                                           
				
				\\ \hline
				\multicolumn{1}{|c|}{\begin{tabular}[c]{@{}c@{}}\end{tabular}}       & \multicolumn{1}{c|}{Supervised}  & \multicolumn{1}{c|}{\begin{tabular}[c]{@{}c@{}}SVM with\\ autoencoder\end{tabular}} & \multicolumn{1}{c|}{\begin{tabular}[c]{@{}c@{}}U.of California\\Merced\end{tabular}} & \multicolumn{1}{c|}{90\%}                                                           
				
				\\ \hline
				\multicolumn{1}{|c|}{}       & \multicolumn{1}{c|}{Supervised}  & \multicolumn{1}{c|}{\begin{tabular}[c]{@{}c@{}}LSTM\end{tabular}} & \multicolumn{1}{c|}{\begin{tabular}[c]{@{}c@{}}U. of California\\ Merced\end{tabular}} & \multicolumn{1}{c|}{69\%}                                                           
				
				\\ \hline
				\multicolumn{1}{|c|}{{[}33{]}}       & \multicolumn{1}{c|}{Traditional}   & \multicolumn{1}{c|}{\begin{tabular}[c]{@{}c@{}}Fast Fourier \\ Transform envelop\end{tabular}}                     & \multicolumn{1}{c|}{\begin{tabular}[c]{@{}c@{}}U.of New \\ South Wales\end{tabular}}      & \multicolumn{1}{c|}{Kurtosis}                                                                  \\ \hline
			\end{tabular}
		\end{adjustbox}
	\end{table}
	
	\subsection{Results discussion}
	Comparing the FaultFace methodology proposed in this paper with the methods in Table.\ref{BBComparison}, an accuracy of 100\% can be reached using the proper FacePortrait as well as the DCGAN network for dataset balancing.  Notice that most of the methods listed on table \ref{BBComparison}  requires a previous stage of feature extraction using different techniques, in order to create a rich training feature dataset to improve the detection accuracy. In the case of FaultFace method, automatic feature extraction is performed due to the use of trained CNN networks for the fault detection tasks. However, the quality of the balanced dataset is relevant for the success of the methodology. It can be observed when the DCGAN is replaced with a GAN network for dataset balancing, the accuracy of the detection is reduced as shown in Table \ref{DCGANConfPerf} and Table \ref{GANConfPerf}. A possible cause for this condition is because DCGAN incorporate convolutional layers that can be trained for specific feature extraction and generation. But, in the case of GAN networks, classic multilayer perceptron layers are employed, which requires more training time and number of hidden elements to produce the desired data.
	
	In the case of LSTM network, considering that the sampling frequency of the vibration signals in \cite{b26} is too high, more cell may be required to improve the method detection as well as different minibatch size to reduce the need of padding operators that affect the detection quality. For the SVM with autoencoder fault detection algorithm, the balanced dataset generated either with GAN or DCGAN networks improve significantly the overall performance of the detection over the unbalanced dataset.
	
	So that, the combination between automated feature extraction layers, the dataset balancing methods (DCGAN), and deep learning classification algorithms as CNN makes a siginificative difference performing fault detection for ball bearing elements regarding to other methodologies. For these reasons, we can conclude that FaultFace is a suitable methodology for failure detection for unbalanced datasets that could be employed not only for ball bearing joints but also for different industrial processes.
	
	\section{Conclusions}
	This paper presented the FaultFace method for failure detection on Ball-Bearing joints based on DCGAN and CNN networks. The proposed method uses a FacePortrait, a 2D representation of a signal that can be obtained using time-frequency representations. For this system, six different FacePortraits were employed using CWT, CMR, Haar, Hankel, Gram, and Toeplitz transformations for six operating conditions composed of the nominal operation and five failure behaviors. A DCGAN network was trained to generate new FacePortraits based on the available data of nominal and failure behaviors to produce a balanced dataset that improves failure detection performance. The balanced dataset of face portraits is employed to train a CNN network that classifies between nominal and failure behaviors. The CNN validation is performed employing the original dataset of the ball bearing system. The FaultFace methodology is also performed using a GAN instead of the DCGAN network. Besides, an LSTM and SVM with autoencoder networks were trained to be compared with the Faultface methodology. Obtained results show that using the CWT, CWT, Hankel, Gram, and Toep face portraits of the vibration signals, the FaultFace methodology performs an accurate detection of nominal and failure behavior. However, the Haar FacePortrait has a reduced accuracy due to the absence of recognizable features in this representation. Also, when GAN is employed with the FaultFace methodology, the quality of the balanced dataset is different, reducing the FaultFace method accuracy. Likewise, using the balanced dataset produced by GAN and DCGAN networks shows an important improvement for the SVM with autoencoder detection algorithm. Also, a comparison between the FaultFace with other fault detection methods for the ball bearing system shows that the FaultFace offer excellent accuracy without the need to perform additional feature extraction and dimensional data reduction. Thus, it is possible to say that the FaultFace method can be considered as an alternative for failure detection not only for the Ball-Bearing problem but also for different industrial processes with unbalanced datasets and complex dynamics. As future works, the real-time implementation of the FaultFace methodology using edge computing devices is proposed as well as the extension of this methodology to other industrial processes than ball bearing elements. Moreover, the development of compressive deep learning algorithms is proposed to perform deep neural stable control techniques that introduce cognitive capabilities on the edge to smart industrial processes monitoring, prognosis, and control.
	
	\bibliography{mybibfile}

\end{document}